\documentclass[aps,prb,twocolumn,floats,showpacs]{revtex4}

\usepackage{epsfig}

\usepackage{amsmath}

\usepackage{amssymb}

\usepackage{bm}
\usepackage{color}

\def\up{\uparrow}
\def\down{\downarrow}

\def\veps{\varepsilon}
\newcommand{\ra}{\rangle}
\newcommand{\la}{\langle}

\begin{document}

\title{SU(12)  Kondo Effect in Carbon Nanotube Quantum Dot}

\author{Igor Kuzmenko and Yshai Avishai}
\affiliation{Department of Physics, Ben-Gurion University of
  the Negev Beer-Sheva, Israel }

\date{\today}

\begin{abstract}
We study the Kondo effect in a CNT(left lead)-CNT(QD)-CNT(right
lead) structure. Here CNT is a single-wall metallic carbon
nanotube, for which 1) the valence and conduction bands of
electrons with zero orbital angular  momentum ($m=0$) coalesc at
the two valley points ${\bf{K}}$ and ${\bf{K}}'$ of the first
Brillouin zone and 2) the energy spectrum of electrons with  $m
\ne 0$ has a gap whose size is proportional to $|m|$. Following
adsorption of hydrogen atoms and application of an appropriately
designed gate potential, electron energy levels in the CNT(QD) are
tunable to have: 1) two-fold spin degeneracy; 2) two-fold isospin
(valley) degeneracy; 3) three-fold orbital degeneracy $m=0,\pm1$.
As a result, an SU(12) Kondo effect is realized with remarkably
high Kondo temperature.  Unlike the SU(2) case, the low temperature
conductance and magnetic susceptibility have a peak at finite temperature. Moreover, the magnetic
susceptibilities for parallel and perpendicular magnetic fields (WRT the tube axis)
display anisotropy with a universal ratio $\chi_{\rm{imp}}^\parallel /
\chi_{\rm{imp}}^\perp=\eta$ that depends only on the electron's orbital and
spin $g$ factors.
\end{abstract}


\maketitle

\section{Introduction}

\noindent \textbf{Background:}
Kondo tunnelling through carbon nanotube quantum dots CNT(QD) has
recently become a subject of intense theoretical\cite{CN-QD-prb11,%
CN-QD-prb10, CN-QD-prb09, CN-QD-prb06, Martins-PRB07,%
Martins-JPhys09} and experimental\cite{Cobden, CN-QD-prl12,%
CN-QD-Nanotech11, CN-QD-prb07, CN-QD-Nature05, CN-QD-prb04}
studies. One of the motivations for pursuing this research
direction is the quest for achieving an exotic Kondo effect with
SU(N) dynamical symmetry,\cite{Dynsym, Kikoin-Avishai-02, KKA04,%
KKA-prl06, KKA-prb06} based on the peculiar properties of electron
spectrum in CNT.\cite{Ando-93-1,Ando3} Achieving SU(4) symmetry is
natural because the energy spectrum of metallic CNT consists of
two independent valleys that touch at the ${\bf{K}}$ and
${\bf{K}}'$ points of the Brillouin zone. The energy levels
possess degeneracy in both spin ($\up,\down$) and isospin (or
valley ${\bf{K}},{\bf{K}}'$) quantum numbers. Thus, due to both
spin and isospin degeneracy, an SU(4) Kondo effect takes place.%
\cite{CN-QD-prb11, CN-QD-prb10, CN-QD-prb09, CN-QD-prb07,%
CN-QD-Nature05,CN-QD-prb04}

\noindent \textbf{Motivation:} Achieving even higher degeneracy
SU(N$>$4) of the QD is highly desirable. Firstly, the Kondo
temperature dramatically increases with $N$. Secondly, there is a
hope to expose novel physical observables that are peculiar to
these higher symmetries. In the present device, higher degeneracy
may be obtained by employing the orbital (cylindrical) symmetry of
electron states in CNT, an option which so far has not been
effectively employed in this quest. In order to manipulate these
orbital features, we use the fact that adsorption of oxygen,
hydrogen or fluorine atoms on the surface of the CNT gives rise to
gap opening in the spectrum of the metallic CNT.%
\cite{H2O-graphene-JETP-Lett, H2O-graphene-JPhys-12} Realization
of SU(N$>4$) Kondo effect then becomes feasible, since there is
now spin, isospin (valley) and orbital degeneracy.

\noindent \textbf{The main objectives:} The main goals of the
present work are: 1) To show that SU(12) Kondo effect in the
CNT(left lead)-CNT(QD)-CNT(right lead)  structure is indeed
achievable and 2) To elucidate the physical content of this
structure at the Kondo regime as encoded by tunneling conductance
and the magnetic susceptibility. The first goal obtains by
designing the electron spectrum in the CNT(QD) to have a 12-fold
degeneracy following adsorption of hydrogen atoms combined with an
application of  a non-uniform gate potential. Namely, the energy
levels of the central element CNT(QD) are tunable into a
three-fold orbital degeneracy for $m=0,\pm1$ (where $m$ is the
component of the orbital angular momentum along the CNT axis). The
second goal is achieved through quantitative analysis, based on
perturbation theory at high temperatures and mean field slave
boson formalism at low temperature.

\begin{figure}[htb]
\includegraphics[width=70 mm,angle=0]
   {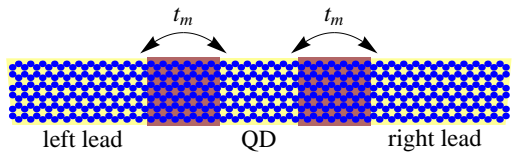}
 \caption{\footnotesize
   (Color online) CNT(left lead)-CNT(QD)-CNT(right lead)  junction.}
  \label{Fig-CN-QD-CN}
\end{figure}

\noindent \textbf{The main results:} The energy spectrum of the
CNT(QD) gated by a spacially modulated potential is elucidated,
and the possibility to get a CNT(QD) with twelve-fold degenerate
quantum states is substantiated. This CNT(QD) is then integrated
into a tunnelling junction CNT(left lead)-CNT(QD)-CNT(right lead)
as shown in Fig.~\ref{Fig-CN-QD-CN}. When the ground state of the
interacting CNT(QD) is occupied by a single electron, Kondo
tunneling with SU(12) dynamical symmetry is realized. This exotic
Kondo effect is quantitatively analyzed. First, the corresponding
Kondo temperature is calculated and shown to be much higher than
in the standard SU(2) Kondo effect. The tunneling conductance
$G(T)$ and the magnetic susceptibilities $\chi_{\rm{imp}}^{\parallel}(T)$,
$\chi_{\rm{imp}}^{\perp}(T)$ for respective magnetic fields parallel and
perpendicular to the CNT axis are calculated in the weak
($T\gg{T_K}$) and strong ($T<T_K$) coupling regimes.

The low temperature dependencies  of both $G(T)$ and
$\chi(T)$ are entirely distinct from their analogs pertaining to the SU(2)
Kondo effect in quantum dot. \cite{KEQD}
More concretely, the temperature
dependence of both quantities is shown to have a peak at {\em finite temperature},
unlike the familiar monotonic behavior encountered in the ordinary SU(2) Kondo effect
in quantum dots.
Moreover, inspection of the magnetic susceptibility
exposes an observable peculiar to the SU(12) symmetry (or other
SU(N) symmetry with $N>2$): It is shown that the
magnetic response is anisotropic, that is, $\chi_{\rm{imp}}^{\parallel}(T) \ne
\chi_{\rm{imp}}^{\perp}(T)$. Even more remarkable, the ratio $\eta \equiv
\chi_{\rm{imp}}^{\parallel}(T)/ \chi_{\rm{imp}}^{\perp}(T)$ is
a ``universal number" depending only on
$g_{\mathrm{orb}}$ and $g_{\mathrm{spin}}$, that are the orbital
and spin $g$ factors of electrons in the CNT(QD) (and not on
temperature).

These distinctions open the door for experimental manifestation of this peculiar junction.
This is helped by the unusually high Kondo temperature that enables the
measurement of the tunneling conductance in the Kondo regime at
relatively high temperature.

%---------------- CN-QD-CN ---------------------------

\noindent \textbf{Organization}: This paper is structured as
follows: In Section \ref{sec-model}, we describe the basic
structure of an infinitely long metallic CNT with adsorbed
hydrogen atoms. The energy spectrum of the CNT(QD) is discussed in
Section \ref{sec-dot}. The Anderson model for the tunnel junction
is introduced in Section \ref{sec-Anderson}, followed by Section
\ref{sec-Kondo} in which the Anderson Hamiltonian in the local
moment regime is mapped on a spin Hamiltonian, poor-man scaling
equations for the coupling constants are derived, and the Kondo
temperature is evaluated. In Sections \ref{sec-cond} and Section
\ref{sec-magn} the results of our  calculations of the tunnelling
conductance and the magnetic susceptibilities are respectively
presented both in the weak and strong coupling regimes. The main
achievements of the present work are summarized in Section
\ref{sec-concl}. Analysis of the electron wave functions in the
CNT(QD) with adsorbed hydrogen atoms under the appropriate gate
potential is relegated to Appendix \ref{append-dot}. Zeeman
splitting for electrons in CNT subject to an external magnetic
field is calculated in Appendix \ref{append-Zeeman}.
 The ratio
$\chi_{\rm{imp}}^{\parallel}(T)/ \chi_{\rm{imp}}^{\perp}(T)$
is derived in Appendix \ref{append-fluct-dissip} using the
fluctuation-dissipation theorem.

\section{Model}
  \label{sec-model}

Characteristic energy dispersion relation for electron in CNT is
derivable from the special band structure of a graphene
sheet.\cite{Ando-93-1, Ando3, egger10, Egger} Let
${\bf{c}}_{n_1n_2}$ denote the chiral vector that represents a
possible rolling of graphene into a CNT. When $n_1-n_2$ is an
integer multiple of 3, a CNT becomes a zero-gap semiconductor.
Else, it becomes a semiconducting nanotube with a finite band
gap.\cite{Ando-93-1,Ando3} The band structure of a metallic CNT
exhibits two Dirac points with a right- and left-moving branch
around each Dirac point. A peculiar consequence of the Dirac
nature of charge carriers in CNT is that electrons can tunnel
through a potential barrier  without
reflection.\cite{graphene-thesis} This  Klein paradox prevents a
practical aspect of CNT: it is virtually impossible to trap an
electron in between potential barriers, as it can escape out. It
also hinders the formation of a gap in the band spectrum.
Fortunately, this can be circumvented by chemical modification of
the CNT. In
Refs.\cite{H2O-graphene-JETP-Lett,H2O-graphene-JPhys-12} it is
shown that when radicals such as atomic oxygen, hydrogen or
fluorine are adsorbed on the graphene surface they form covalent
bonds with the carbon atoms. These covalent bonds are realized
since the carbon atoms change their hybridization from $sp^{2}$ to
$sp^{3}$, and that results in the opening of a band gap (similar
to the situation in diamond crystals).  Its size can reach
$2\Delta_g\sim1$~eV depending on the density of adsorbed atoms.%
\cite{H2O-graphene-JETP-Lett,CNT-adsorb-gap-03}

\vspace{6mm}

%---------------- CN-CNT(QD)-CNT ---------------------------
\begin{figure}[htb]
%H=6.28, L=14.65
\centering
\includegraphics[width=70 mm,angle=0]
   {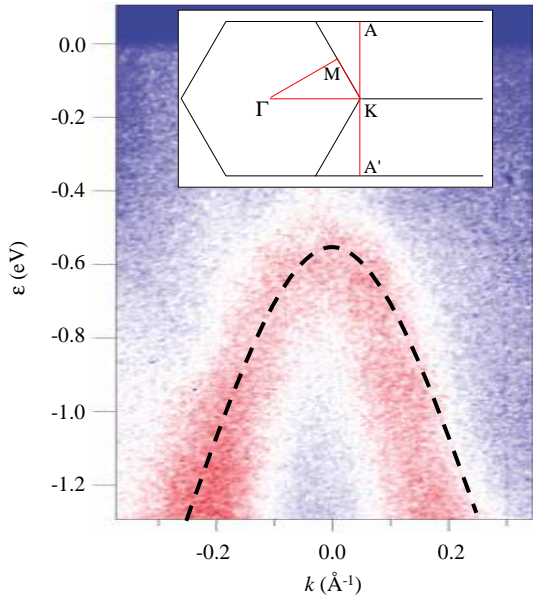}
 \caption{\footnotesize
  (Color online) Observation of a gap opening in hydrogenated
  graphene. Density plot denotes photoemission intensity along
  the $A$-$K$-$A'$ direction of the Brillouin zone (see inset)
  measured in Ref.[\onlinecite{H-on-graphene-nature-10}], whereas
  the dashed line is the spectrum calculated according to
  Eq.(\ref{sp-CNT}). \textbf{Inset}: The Brillouin zone of
  graphene.}
 \label{Fig-sp-graphene-H}
\end{figure}

The energy spectrum of the metallic CNT with adsorbed atoms can
adequately be approximated (at least, at low energy) from that of
a graphene sheet with adsorbed atoms using the formula,
\begin{eqnarray}
  \varepsilon_{k_xk_y} &=&
  \sqrt{(\hbar v k_x)^2+(\hbar v k_y)^2+\Delta_g^2}.
  \label{sp-CNT}
\end{eqnarray}
In the above equation we  keep $k_x$ to be a continuous wave
number for electron motion along the CNT axis and $k_y=m/r_0$ as
discrete wave number for the motion along the circumference
direction.  Here $v$ is the Fermi velocity, $r_0$ is the CNT
radius and the integer $m$ is an orbital quantum number. The
energy spectrum of hydrogenated graphene measured in
Ref.\cite{H-on-graphene-nature-10} is shown in
Fig.\ref{Fig-sp-graphene-H}. It is seen that Eq.(\ref{sp-CNT})
(dashed line) agrees well with experimental data.

\begin{figure}[htb]
%H=6.28, L=14.65
\centering
\includegraphics[width=65 mm,angle=0]
   {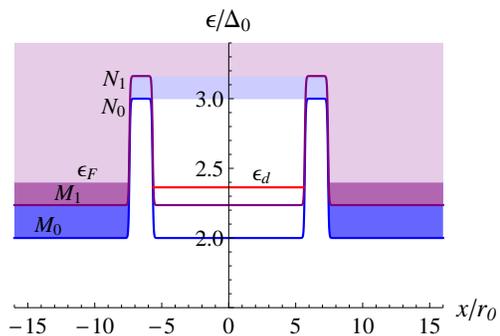}
 \caption{\footnotesize
  (Color online) Left and right tunnel barriers separating left
  and right leads from the quantum dot. Here $M_0=2\Delta_0$,
  $M_1=\sqrt{3}\Delta_0$, $N_0=3\Delta_0$ and
  $N_1=\sqrt{10}\Delta_0$. The twelve fold degenerate
  level is $\varepsilon_d=2.3635\Delta_0$ (red line) and
  the Fermi level is $\epsilon_F=2.4013\Delta_0$. The QD
  half-length is $h=5.6826r_0$ and the barrier width is
  $a=1.73r_0$.}
 \label{Fig-model}
\end{figure}

With present experimental facilities, the density of adsorbed
atoms can be manipulated to be dependent on $x$ in such a way that
the gap $\Delta_g$ is approximately given by the following
function of $x$,
\begin{eqnarray}
  \Delta_g(x) &=&
  \left\{
  \begin{array}{ccl}
  M_0, &{\rm{if}}& |x|<h ~{\rm or}~ |x|>h+a,
  \\
  N_0, &{\rm{if}}& h<|x|<h+a,
  \end{array}
  \right.
  \label{Delta(x)-def}
\end{eqnarray}
where $N_0>M_0$. The Fermi level $\epsilon_F$ is tuned to satisfy
the inequality $N_0>\epsilon_F>M_1$. Thereby, the CNT is divided
into five intervals numbered 1-5, with the following respective
electronic properties: Two intervals (1 and 5), with $|x|>h+a$,
serve as left and right metallic leads.  Two insulating intervals
(2 and 4), with $h<|x|<h+a$, serve as left and right tunneling
barriers. Finally, interval 3 with $|x|<h$ serves as quantum dot
(see Fig.\ref{Fig-model}).

%---------------- CN-chiral ---------------------------

%%%%%%%%%%%%%%%%%%%%%%%%%%%%%%
\section{Energy Levels of CNT(QD)}
  \label{sec-dot}

\begin{widetext}

%---------------- CN-CNT(QD)-CNT ---------------------------
\begin{figure}[htb]
%H=6.28, L=14.65
\centering
 \begin{tabular}{|l|l|}
 \hline
 ~ (a) & ~ (b)
 \\
 ~
 \includegraphics[height=45 mm,angle=0]
    {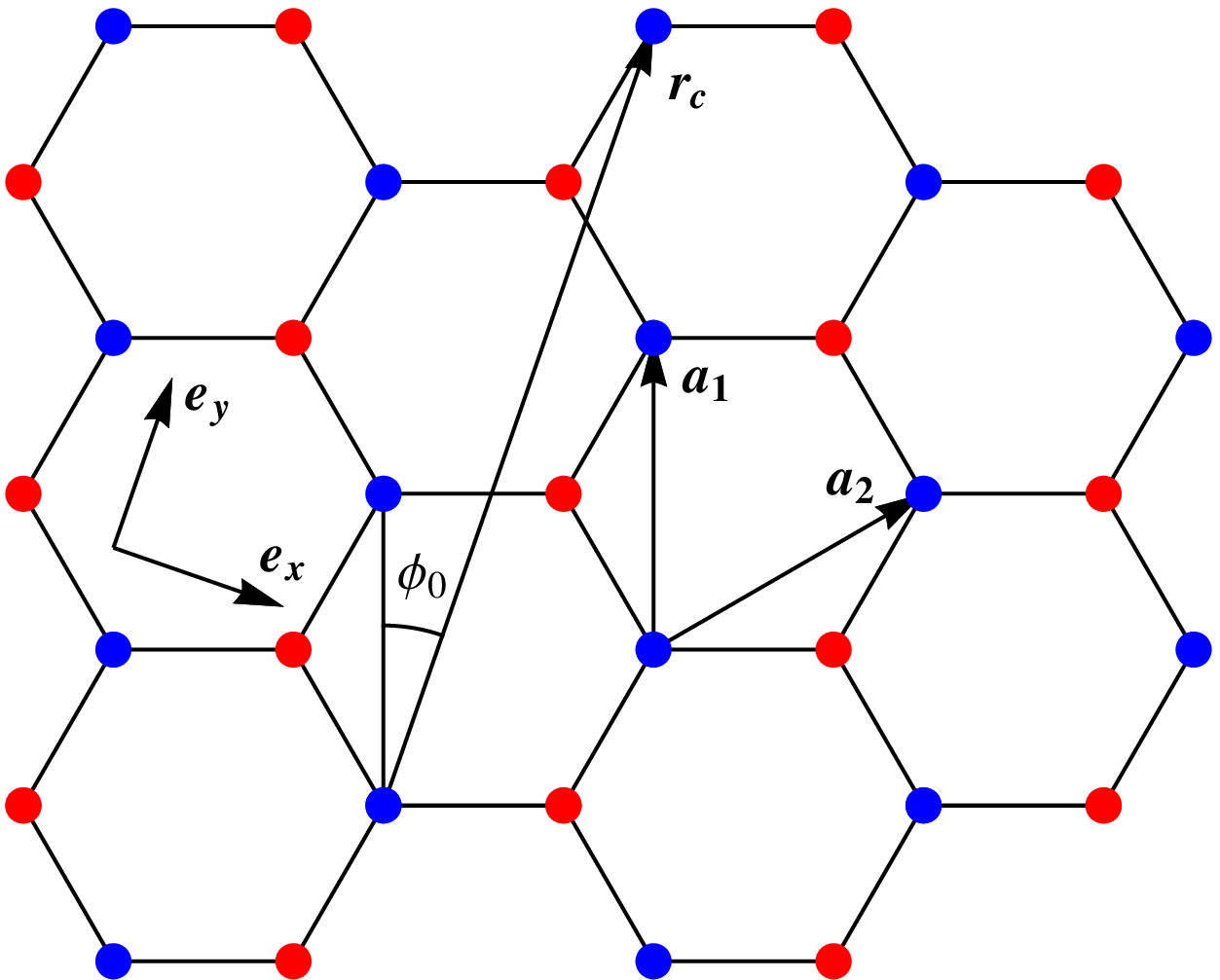}
 ~
 &
 ~
 \includegraphics[height=45 mm,angle=0]
    {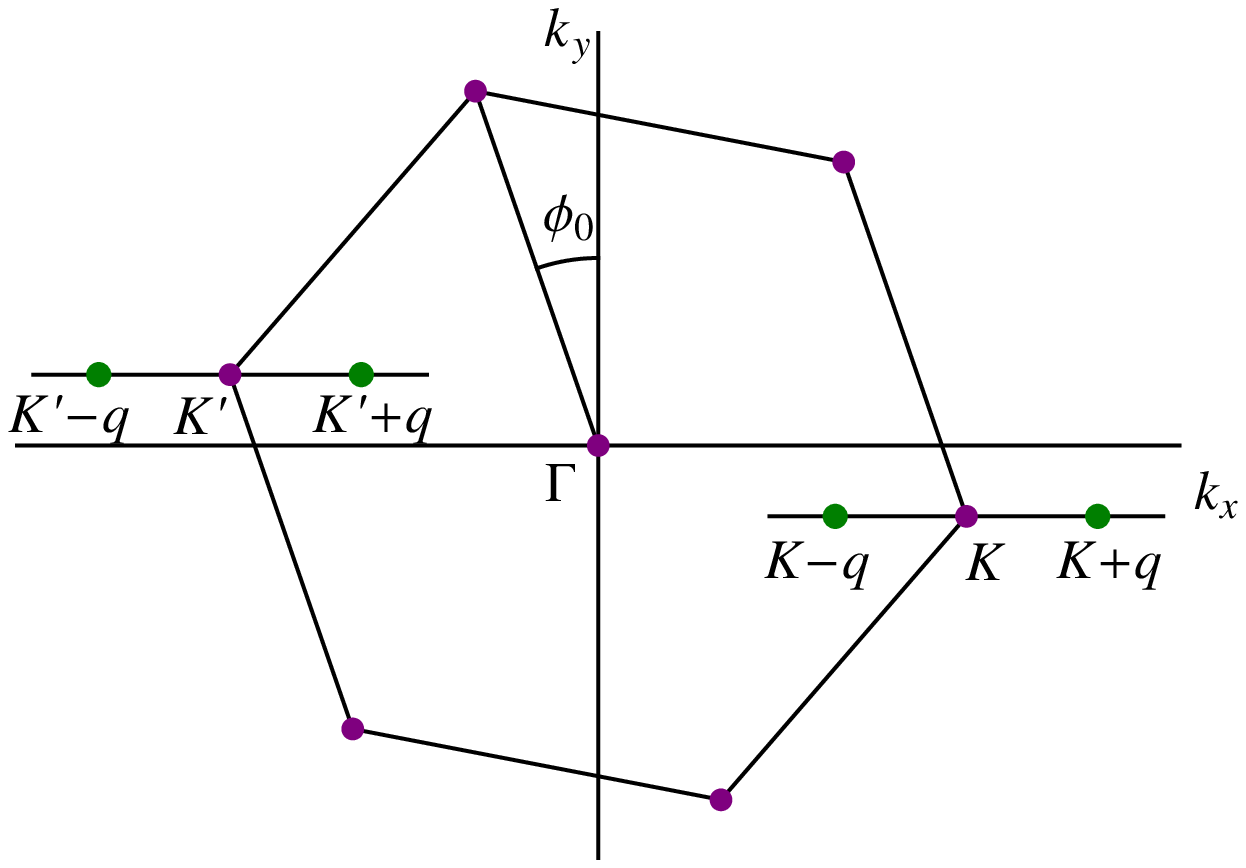}
 ~
 \\
 \hline
 \end{tabular}
 \caption{\footnotesize
  (Color online)
  {\bf{Panel (a)}}: A monoatomic layer of graphene. The red and
  blue dots denote carbon atoms of the sub-lattice A and B. The
  primitive vectors of graphene are ${\bf{a}}_1$ and ${\bf{a}}_2$.
  The nanotube is obtain by choosing the chiral vector
  ${\bf{c}}_{n_1n_2}$, equation (\ref{chiral-vector-def}). The
  unit vectors ${\bf{e}}_x$ and ${\bf{e}}_y$ are fixed in the CNT
  in such a way that ${\bf{e}}_x$ is along the CNT axis, and
  ${\bf{e}}_y$ is along the circumferential direction
  ${\bf{c}}_{n_1n_2}$. The chiral angle between ${\bf{a}}_1$ and
  ${\bf{c}}_{n_1n_2}$ is $\phi_0$.
  {\bf{Panel (b)}}: The first Brillouin zone of graphene.
  $k_x$ is the component of the 2D wave vector ${\mathbf{k}}$ alon
  the CNT axis and $k_y$ is the component of ${\bf{k}}$ in
  the circumferential direction. The angle between ${\bf{K}}$ and
  the axis $k_x$ is $\phi_0-\frac{\pi}{6}$. ${\bf{K}}\pm{\bf{q}}$
  and ${\bf{K}}'\pm{\bf{q}}$ (green dots) are degenerate quantum
  states in the valleys ${\bf{K}}$ and ${\bf{K}}'$.}
 \label{Fig-CNT-chiral}
\end{figure}

\end{widetext}

We describe the quantum states of electrons in the CNT in the
long-wave ${\bf{k}}\cdot{\bf{p}}$ approximation. This
approximation is good when the wave vector ${\bf{k}}$ of the
electron is close to the ${\bf{K}}$ or ${\bf{K}}'$ point of the
first Brillouin zone (BZ) of the hexagonal lattice of the CNT,
i.e., when $|{\bf{k}}-{\bf{K}}|\ll{K}$ or
$|{\bf{k}}-{\bf{K}}'|\ll{K}$ [see Figure \ref{Fig-CNT-chiral} for
illustration]. However, when electron is rejected from the edges
of the CNT QD, the possible transitions between the valleys
${\bf{K}}$ and ${\bf{K}}'$ cannot be described in the framework of
long-wave approximation. Therefore, it will be useful to start our
discussions from the microscopic tight-binding
model.\cite{Ando-93-1,Ando3}

A CNT is specified by a chiral vector
\begin{eqnarray}
  {\bf{c}}_{n_1n_2}=n_1{\bf{a}}_1+n_2{\bf{a}}_2,
  \label{chiral-vector-def}
\end{eqnarray}
where ${\bf{a}}_1$ and ${\bf{a}}_2$ are the basis vectors
[$|{\bf{a}}_1|=|{\bf{a}}_2|=a_0=2.46$~{\AA}], $n_1$ and $n_2$ are
integers. A CNT is obtained by rolling a 2D graphene sheet such
that the atom at the origin coincides with the atom at
${\bf{c}}_{n_1n_2}$. Then $|{\bf{c}}_{n_1n_2}|=2\pi{r_0}$ is the
length of the CNT circumference and $r_0$ is the CNT radius. We
specify the CNT by a chiral angle $\phi_0$, the angle between
${\bf{c}}_{n_1n_2}$ and the basis vector ${\bf{a}}_1$, as shown in
Figure \ref{Fig-CNT-chiral}. The hexagonal symmetry of grapheene
gives us the condition $-\frac{\pi}{6}<\phi_0\le\frac{\pi}{6}$.
Two special values of $\phi_0$ are $\phi_0=0$ and
$\phi_0=\tfrac{\pi}{6}$. For $\phi_0=0$, a zigzag CNT is
constructed, while for $\phi=\frac{\pi}{6}$, one has an armchair
CNT.\cite{Ando-93-1, Ando3}

When an electron is scattered off an effective potential given in
Eq.~(\ref{Delta(x)-def}), the component $k_x$ of the 2D wave
vector ${\bf{k}}$ is not a good quantum number, whereas $k_y$ is
still a good quantum number. As a result, for most types of
nanotubes with $\phi_0 \ne \tfrac{\pi}{6}$ [that is, except
armchair ones] the vectors ${\bf{K}}$ and ${\bf{K}}'$ are not
collinear to the CNT axis [see Figure \ref{Fig-CNT-chiral}b], and
therefore the electron that is localized by the potential
(\ref{Delta(x)-def}) can change its wave vector from
${\bf{K}}+{\bf{q}}$ to ${\bf{K}}-{\bf{q}}$ or from
${\bf{K}}'+{\bf{q}}$ to ${\bf{K}}'-{\bf{q}}$, and there is no
quantum transitions between the valleys ${\bf{K}}$ and
${\bf{K}}'$. For an armchair CNT, the vectors ${\bf{K}}$ and
${\bf{K}}'$ are collinear with the axis of the CNT, and therefore
there are quantum transitions from ${\bf{K}}$ to ${\bf{K}}'$ which
lift the inter-valley degeneracy. In what follows, we will
consider the CNT QD's which possess the inter-valley degeneracy
(that is, $\phi_0 \ne \tfrac{\pi}{6}$).

When $|{\bf{q}}|\ll{K}$, the single electron wave functions and
the corresponding energy spectrum of the CNT(QD) are deduced from
the corresponding analog of the Dirac equation which in the
present geometry takes the form,
\begin{eqnarray}
  \tilde{H}_{d}
  \Phi_{mn}(x,\phi)
  &=&
  \epsilon
  \Phi_{mn}(x,\phi).
  \label{Dirac-eq-dot}
\end{eqnarray}
Here $\Phi_{mn}(x,\phi)$ is the wave function with principal
quantum number $n$ [$n=1,2,3,\ldots$] and magnetic quantum number
$m$ [$m=0,\pm1,\pm2,\ldots$]. The Hamiltonian of the QD in the
${\bf{k}}\cdot{\bf{p}}$ approximation is
$\tilde{H}_{d}={\bf{d}}(x)\cdot\boldsymbol\tau$, where
\begin{eqnarray*}
  &&
  {\bf{d}}(x)=
  {\bf{d}}_{\rm{in}}
  \vartheta(h-|x|)+
  {\bf{d}}_{\rm{out}}
  \vartheta(|x|-h),
  \nonumber \\ &&
  {\bf{d}}_{\rm{in}} =\hbar v
  \Big(
      {\mathbf{e}}_x
      k_x+
      {\mathbf{e}}_y
      k_y
  \Big)+
  {\mathbf{e}}_z
  M_0,
  \nonumber \\ &&
  {\bf{d}}_{\rm{out}} =
  \hbar v
  \Big(
      {\mathbf{e}}_x
      k_x+
      {\mathbf{e}}_y
      k_y
  \Big)+
  {\mathbf{e}}_z
  N_0.
\end{eqnarray*}
Here $\boldsymbol\tau=(\tau_x,\tau_y,\tau_z)$ is the vector of
Pauli matrices acting in the iso-spin space, $\vartheta(x)$ is the
step function and
\begin{eqnarray}
  &&
  k_x = -i\partial_x,
  \ \ \ \ \
  k_y = -\frac{i}{r_0}~\partial_{\phi}.
\end{eqnarray}
The function $\Phi_{mn}(x)$ is calculated in Appendix
\ref{append-dot}. The energy levels of the QD are obtained by
solving the equation,
\begin{eqnarray}
  {\cal{F}}_{\down}(\epsilon)~
  \sqrt{\epsilon+M_m}~
  \cos
  \Big(
      k_x h+
      \frac{n\pi}{2}
  \Big)
  = ~~~~~
  \nonumber \\ =
  {\cal{F}}_{\up}(\epsilon)~
  \sqrt{\epsilon-M_m}~
  \sin
  \Big(
      k_x h+
      \frac{n\pi}{2}
  \Big),
  \label{Eq-for-spectrum}
\end{eqnarray}
where
\begin{eqnarray}
  {\cal{F}}_{\sigma}(\epsilon) &=&
  \sum_{\sigma'}
  \frac{\sqrt{\epsilon(N_m+\sigma'\epsilon)}}
       {\sqrt{2N_m}}~
  \big(\chi^{(1)}_{\sigma}\cdot\chi^{(2)}_{\sigma'}\big).
  \label{cal-F-def}
\end{eqnarray}
$\chi_{\sigma}^{(\nu)}$ [$\nu=1,2$] are eigenspinors of the
operators
$$
  \hat{M}_m^{\nu}=
  m\Delta_0\tau_y+
  M_0^{\nu}\tau_z,
$$
$M_m^{1}=M_m$, $M_m^{2}=N_m$. Explicitly,
\begin{eqnarray}
  \chi_{\up}^{(\nu)} =
  \frac{1}
       {\sqrt{2M_m^{\nu}
        (M_m^{\nu}+M_0^{\nu})}}~
  \left(
  \begin{array}{c}
    M_m^{\nu}+M_0^{\nu}
    \\
    i m \Delta_0
  \end{array}
  \right),
  \label{chi-1-2-up-def}
\end{eqnarray}
$\chi_{\down}^{(\nu)}=\tau^x\chi_{\up}^{(\nu)}$.

\begin{figure}[htb]
%H=6.28, L=14.65
\centering
\includegraphics[width=60 mm,angle=0]
   {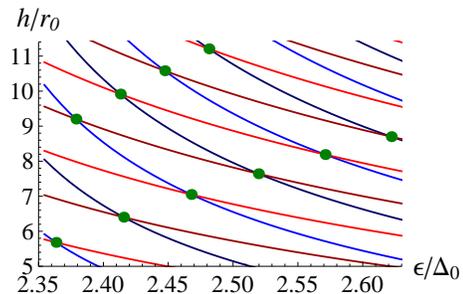}
 \caption{\footnotesize
   (Color online) Parametric diagram $\epsilon$-$h$ describing
   energies $\epsilon$ of the discrete levels of the quantum dot
   for $m=0$ (light red and dark red curves) and $m=\pm1$ (light
   blue and dark blue curves) and different values of the CNT(QD)
   half-length $h$. The crossing points (green dots) denote
   energies with three-fold orbital degeneracy for which SU(12)
   symmetry is expected. Here we use $M_0=2\Delta_0$ and
   $N_0=3\Delta_0$. The light red and light blue curves correspond
   to even principal quantum number $n$ and dark red and dark blue curves
   correspond to odd $n$.
   }
 \label{Fig-Sp-SU12}
\end{figure}

The energy spectrum of the QD for different values of the dot
length $2h$ is shown in Fig.\ref{Fig-Sp-SU12}. The red or blue
curves denote energy levels for $m=0$ or $m=\pm 1$ and different
spatial parities. The light red and light blue curve describe
quantum states with even principal quantum number $n$ [the wave
functions of such states are symmetric with respect to the
inversion ${x}\to{-x}$], whereas the dark red and dark blue curves
correspond to odd $n$. The level crossing points (green dots) are
three fold (orbital) degenerate. At these points, SU(12) symmetry
is expected.

%----------- m = 0 and m = 1 crossing points --------------

\section{Anderson Model}
  \label{sec-Anderson}

We now consider the tunnel junction consisting of  left and right
CNT metallic leads (CNT), and a CNT quantum dot (QD), as shown in
Fig. \ref{Fig-CN-QD-CN}. The Anderson Hamiltonian of the CNT --
CNT (QD) -- CNT junction has the form,
\begin{eqnarray}
  H &=&
  H_0+H_t,
  \ \ \ \ \
  H_0 = H_l+H_r+H_d,
  \label{H-Anderson-def}
  \\
  H_{\alpha} &=&
  \sum_{k\lambda}
  \epsilon_{km}~
  c_{\alpha k \lambda}^{\dag}
  c_{\alpha k \lambda},
  \ \
  \alpha=l,r,
  \label{H-Anders-LR}
  \\
  H_d &=&
  \epsilon_d
  \sum_{\lambda}
  d_{\lambda}^{\dag}
  d_{\lambda}+
  U_d \hat{N}_d(\hat{N}_d-1),
  \label{H-Anders-dot}
  \\
  H_t &=&
  \sum_{\alpha \lambda}
  t_{m}
  \Big\{
      \psi_{\alpha \lambda}^{\dag}
      d_{\lambda}+
      \text{H.c.}
  \Big\},
  \label{H-Anders-T}
\end{eqnarray}
where
$$
  \hat{N}_d =
  \sum_{\lambda}
  d_{\lambda}^{\dag}
  d_{\lambda}.
$$
Here $\lambda=\{\xi,m,\sigma\}$, where
$\xi={\mathbf{K}},{\mathbf{K}}'$ (the isospin) corresponds to
electrons with wave vectors near the ${\bf{K}}$ and ${\bf{K}}'$
corner points in the 2D Brillouin zone, $m$ is the magnetic
quantum number and $\sigma$ is the spin. Finally,
$\psi_{\alpha\lambda}\equiv\psi_{\alpha\lambda}(x=0)$ is a
field operator at $x=0$,
\begin{eqnarray*}
  \psi_{\alpha\lambda}(x) &=&
  \frac{1}{\sqrt{L_{\rm{cnt}}}}
  \sum_{k}
  c_{\alpha k \lambda}
  e^{ikx},
\end{eqnarray*}
$L_{\rm{cnt}}$ is the length of the CNT lead. The tunneling rates
$t_m$ are estimated as
\begin{eqnarray*}
  t_m &\cong&
  \frac{\hbar v}{\sqrt{h}}~
  \frac{M_m}{\epsilon_F}~
  \exp
  \bigg\{
      -\frac{a}{\hbar v}~
       \sqrt{N_m^2-\epsilon_F^2}
  \bigg\}.
\end{eqnarray*}
We choose the parameters $N_0$, $\epsilon_F$ and $a$ such that the
resonance width,
\begin{equation} \label{width}
  \Gamma=4\pi{t}_{m}^2\rho_{m}(\epsilon_F),
\end{equation}
does not depend on $m$. Here $\rho_{m}(\epsilon)$ is the density
of states of electrons with magnetic quantum number $m$,
\begin{equation} \label{DOSm}
  \rho_{m}(\epsilon)=
  \frac{|\epsilon|~
        \vartheta\big(|\epsilon|-M_m\big)}
       {2\pi\hbar v
        \sqrt{\epsilon^2-M_m^2}}.
\end{equation}

\section{Spin Hamiltonian, Scaling Equations and Kondo Temperature}
  \label{sec-Kondo}

The properly tuned CNT(QD) in its ground state has one electron
whose energy $\varepsilon_{0}$ is twelve-fold degenerate
($m=0,\pm1$, $\xi={\bf{K}},{\bf{K}}'$ and $\sigma=\up,\down$).
Tunneling of electrons between the CNT(QD) and the CNT leads,
encoded by $H_t$, Eq. (\ref{H-Anders-T}), changes the number of
electrons in the dot. In the local moment regime, the
Schrieffer-Wolff transformation is then
used\cite{Schrieffer-Wolff-66,Hewson-book} to project out zero and
two electron states ($|0\rangle$ and $|\lambda\lambda'\rangle$.
It maps the Hamiltonian $H$, Eq. (\ref{H-Anderson-def}) onto an effective
Hamiltonian $\tilde{H}=H_l+H_r+H_K$.  Here $H_K$, the  Coqblin-Shrieffer
spin Hamiltonian with the dot states $|0\rangle$ and
$|\lambda\lambda'\rangle$ frozen
out, has the following form,\cite{Hewson-book,Coqblin-Shrieffer-69}
\begin{eqnarray}
  &&
  H_K =
  \frac{1}{24}
  \sum_{\alpha\alpha'}
  \sum_{\lambda}
  K_{mm}~
  \psi_{\alpha'\lambda}^{\dag}
  \psi_{\alpha \lambda}+
  \nonumber
  \\ && ~~~~~ +
  \frac{1}{2}
  \sum_{\alpha\alpha'}
  \sum_{\lambda}
  J_{mm}
  Z^{\lambda\lambda}
  \psi_{\alpha'\lambda}^{\dag}
  \psi_{\alpha \lambda}+
  \label{HK-SU12-def}
  \\ && ~~~~~ +
  \frac{1}{2}
  \sum_{\alpha\alpha'}
  \sum_{\lambda\neq\lambda'}
  J_{mm'}
  X^{\lambda\lambda'}~
  \psi_{\alpha'\lambda'}^{\dag}
  \psi_{\alpha \lambda},
  \label{HK-SU12-def1}
\end{eqnarray}
where $X^{\lambda\lambda'}=|\lambda\rangle\langle\lambda'|$ are
Hubbard operators coupling different degenerate dot states, and
\begin{eqnarray*}
  Z^{\lambda\lambda} &=&
  X^{\lambda\lambda}-
  \frac{1}{N}
  \sum_{\lambda'}
  X^{\lambda'\lambda'},
  \ \ \ \ \
  N=12.
\end{eqnarray*}
The couplings $K_{mm}$ and $J_{mm'}$ are
\begin{eqnarray*}
  J_{mm'} &=&
  J_{mm'}^{(1)}+
  J_{mm'}^{(2)},
  \\
  K_{mm} &=&
  J_{mm'}^{(1)}-
  (N-1)J_{mm'}^{(2)},
  \\
  J_{mm'}^{(1)} &=&
  \frac{2t_{m}t_{m'}}{\epsilon_F-\epsilon_d},
  \\
  J_{mm'}^{(2)} &=&
  \frac{2t_{m}t_{m'}}{U_d-\epsilon_F+\epsilon_d}.
\end{eqnarray*}

Employing the simplifying assumption (\ref{width}) we introduce
the dimensionless coupling constant
\begin{eqnarray} \label{smallj}
  j &=&
  J_{mm'}
  \sqrt{\rho_{m}(\epsilon_F)\rho_{m'}(\epsilon_F)} =
  \nonumber \\ &=&
  \frac{U_d\Gamma}
       {2\pi(\epsilon_F-\epsilon_d)(U_d-\epsilon_F+\epsilon_d)}
  > 0.
\end{eqnarray}
By equation (\ref{width}), $j$ does not depend on the orbital
quantum number $m$ while $J_{mm'}$ and $\rho_m$ do. Within the
standard poor man's scaling technique, the coupling $j(D)$ is
renormalized as the original bandwidth $\bar{D}$ is reduced to
$D<\bar{D}$ by integrating out high energy excitations. Within the
same assumption on $\Gamma$, the constants $K_{mm}$ are not
renormalized and therefore the interaction terms proportional to
$K_{mm}$ can be considered as part of potential scattering.

The scaling equation  for $j(D)$ supported by the initial
condition at $\bar{D}$ reads,
\begin{eqnarray}
  &&
  \frac{\partial j}{\partial\ln{D}}
  =
  -\frac{Nj^2}{2},
  \label{scale-eq-SU12}
  \\
  &&
  j(\bar{D})=
  \frac{U_d\Gamma}
       {2\pi(\epsilon_F-\epsilon_d)
        (U_d-\epsilon_F+\epsilon_d)}.
  \nonumber
\end{eqnarray}

Equation (\ref{scale-eq-SU12}) has the solution
\begin{eqnarray}
  j(T) &=& \frac{2}{N\ln(T/T_K)},
  \label{scale-res-SU12}
\end{eqnarray}
where the Kondo temperature (the scaling invariant of the RG
equation) is given by,
\begin{eqnarray}
  T_K =
  \bar{D}
  \exp
  \bigg[
       -\frac{4\pi
              (\epsilon_F-\epsilon_d)
              (U_d-\epsilon_F+\epsilon_d)}
             {NU_d\Gamma}
  \bigg].
  \label{TK-SU12}
\end{eqnarray}
The argument of the exponent is six time smaller than the one
obtained for SU(2) Kondo effect, implying the
${T_K}$[SU(12)]$\gg{T_K}$[SU(2)].

\section{Conductance}
  \label{sec-cond}
In this section we will calculate the tunneling conductance $G(T)$
of the CNT(left lead)-CNT(QD)-CNT(right lead) junction in the
Kondo regime. The calculation is carried out in the weak and
strong coupling regimes characterized respectively by  $T \gg T_K$
and $T<T_K$. In the weak coupling regime, perturbation RG
formalism is used to calculate the non-linear conductance within
the Keldysh non-equilibrium Green's function formalism. In the
strong coupling regime the mean field slave boson formalism is
employed, which is appropriate only within linear response.\\

%%%%%%%%G(T)%%%%%%%%%%% in weak coupling regime
\noindent \textbf{Conductance in the Weak Coupling Limit:}\\
Calculations of the tunneling conductance in the weak coupling
regime are carried out below using the Keldysh technique in order
to treat a system out of equilibrium. The required quantities to
be used below are the Keldysh electron matrix Green's functions
(GF) $g_{a}$ for $a=lm,rm,f$ standing for left lead, right lead
and dot respectively,
\begin{equation}
  g_{a}=
  \begin{pmatrix}
  g_{a}^{R} & g_{a}^{K}
  \\
  0 & g_{a}^{A}
  \end{pmatrix},
  \label{gK}
\end{equation}
where the superscripts refer to retarded ($R$), advanced ($A$) and
Keldysh ($K$) types of the GF. The explicit expressions are,
\begin{eqnarray}
  &&
  g_{\alpha{m}}^{R}=-g_{\alpha{m}}^{A}=-i\pi\rho_m,
  \nonumber \\ &&
  g_{\alpha{m}}^{K}(\epsilon)=
  -2i\pi\rho_m(1-2f(\epsilon)),
  \label{GF-LR}
  \\
  &&
  g_{f}^{R/A}(\epsilon) =
  \frac{1}{\epsilon-\epsilon_d\pm{i\eta}},
  \nonumber \\ &&
  g_{f}^{K}(\epsilon) =
  -\frac{2i\eta(1-2f(\epsilon))}
        {(\epsilon-\epsilon_d)^2+\eta^2},
  \label{GF-dot}
\end{eqnarray}
where  $f(\epsilon)$ is the Fermi  function. Within the Keldysh
formalism, the tunneling current from the left to the right lead
is
\begin{eqnarray}
  I =
  \frac{i e}{24\hbar}
  \sum_{\lambda}
  K_{mm}
  \Big(
      \psi_{l\lambda}^{\dag}
      \psi_{r\lambda}-
      \psi_{t\lambda}^{\dag}
      \psi_{l\lambda}
  \Big)+
  ~~~~~ ~~ \nonumber
  \\ +
  \frac{i e}{2\hbar}
  \sum_{\lambda}
  J_{mm}~
  Z^{\lambda\lambda}
  \Big(
      \psi_{l\lambda}^{\dag}
      \psi_{r\lambda}-
      \psi_{r\lambda}^{\dag}
      \psi_{l\lambda}
  \Big)+
  \nonumber
  \\ +
  \frac{i e}{2\hbar}
  \sum_{\lambda\ne\lambda'}
  J_{mm'}
  X^{\lambda\lambda'}
  \Big(
      \psi_{l\lambda'}^{\dag}
      \psi_{r\lambda}-
      \psi_{r\lambda'}^{\dag}
      \psi_{l\lambda}
  \Big).
  \label{cur-oper-def}
\end{eqnarray}
In addition to the exchange constant $j$, Eq.~(\ref{smallj}), the
conductance depends also on the dimensionless parameter $k$, defined
as (see comment after Eq.~(\ref{smallj})),
\begin{equation} \label{smallk}
  k=
  K_{mm}
  \rho_m=
  \frac{\Gamma
        \big(2U_d-13\epsilon_F+13\epsilon_d\big)}
       {4\pi
        \big(\epsilon_F-\epsilon_d\big)
        \big(U-\epsilon_F+\epsilon_d\big)}.
\end{equation}
To second order in $j$ and $k$ the conductance
$G=\partial\langle{I}\rangle/\partial{V}$ is,
\begin{eqnarray}
  G_2 &=&
  \frac{\pi e^2}{2N\hbar}~
  \Big(
      k^2+
      (N^2-1)j^2
  \Big),
  \label{G2-SU12}
\end{eqnarray}
while only $j$ contributes to the third order correction to the
conductance,
\begin{eqnarray}
  G_3 =
  \frac{(N^2-1)\pi e^2}{4\hbar}~
  j^3
  \ln\bigg(\frac{\bar{D}}{\sqrt{T^2+(eV)^2}}\bigg).
  \label{G3-SU12}
\end{eqnarray}

Due to the large pre-factor and the logarithmic term, which,
strictly speaking, is not small either, $G_3$ is not small as
compared with $G_2$.  Hence, expansion up to third order in $j$ is
inadequate. Instead, we derive an expression for the conductance
in the leading logarithmic approximation using the RG equations
(\ref{scale-eq-SU12}).

In the following analysis we split the second order contribution
to the conductance, Eq. (\ref{G2-SU12}), in two parts: The first
part results from exchange co-tunneling, which is proportional to
$j^2$, while the second part is due to regular co-tunneling, which
is proportional to $k^2$. The regular co-tunneling contribution
containing $k^2$ does not grow at low temperatures and/or bias,
and therefore it does not contribute to the Kondo effect. The
exchange co-tunneling contains a term $j^2$ which demonstrates
logarithmic enhancement of the conductance at low temperatures
[see Eq. (\ref{scale-res-SU12})] and contributes to the Kondo
effect. Therefore, we single out the exchange contribution in the
second order term,
\begin{eqnarray}
  G_2^{\rm{exch}}(D) &=&
  \frac{(N^2-1)\pi e^2}{2N\hbar}~
  j^2(D).
  \label{G2-SU12-exch}
\end{eqnarray}
The condition imposing invariance of the conductance under ``poor
man's scaling'' transformation has the form,
\begin{eqnarray}
  \frac{\partial}{\partial \ln D}
  \bigg\{
       G_2^{\rm{exch}}(D)+
  ~~~~~ ~~~~~
  ~~~~~ ~~~~~
  ~~~~~ ~~~~~
  ~~~~
  \nonumber \\ +
       \frac{(N^2-1)\pi e^2}{4\hbar}~
       j^3
       \ln\bigg(\frac{D}{\sqrt{T^2+(eV)^2}}\bigg)
  \bigg\}
  = 0.
  \label{dG-dD-SU12}
\end{eqnarray}
Within the accuracy of this equation, when differentiating the
second term, we should neglect any implicit dependence on $D$
through the couplings $j$. Eq. (\ref{dG-dD-SU12}) yields the
scaling equation (\ref{scale-eq-SU12}). The renormalization
procedure should proceed until the bandwidth $D$ is reduced to a
quantity
$$
  d(T,V)=
  \sqrt{(eV)^2+T^2}.$$
At this point, the third order correction to the conductance
vanishes and the current and conductance can be calculated in the
Born approximation, as in Eq. (\ref{G2-SU12-exch}).%
\cite{Kaminski-Nazarov-Glazman-00} The expression for the
conductance for ${\rm{Max}}(T,|eV|){\gtrsim}T_K$ is,
\begin{eqnarray}
  G(T,V) =
  \frac{\pi^2 {\cal{N}}~G_0}
       {\ln^2\big(d(T,V)/T_K\big)},
  \label{G-SU12-low-D}
\end{eqnarray}
where
\begin{eqnarray}
  &&
  {\cal{N}} =
  \frac{2(N^2-1)}{N^3},
  \ \ \ \ \
  G_0 =
  \frac{e^2}{\pi\hbar}.
  \label{G0-def}
\end{eqnarray}

%---------------- G(T)-weak ---------------------------
\begin{figure}[htb]
%H=6.28, L=14.65
\centering
\includegraphics[width=65 mm,angle=0]
   {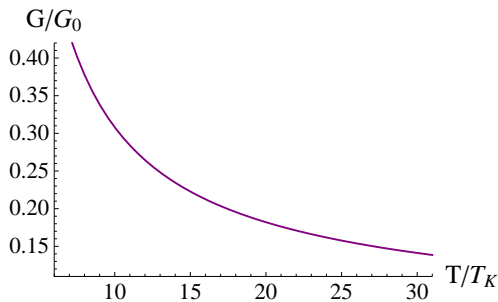}
 \caption{\footnotesize
  (Color online) The zero bias conductance
  (\ref{G-SU12-low-D}) as function of temperature in the weak
  coupling regime.}
  \label{Fig-G-SU12-weak}
\end{figure}

\begin{figure}[htb]
%H=6.28, L=14.65
\centering
\includegraphics[width=65 mm,angle=0]
   {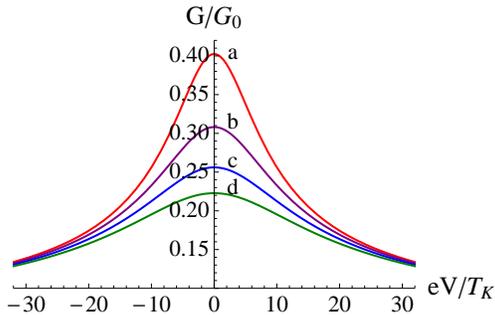}
 \caption{\footnotesize
  (Color online) The nonlinear conductance
  (\ref{G-SU12-low-D}) as function of applied bias in the weak
  coupling regime for $T=7.5T_{K}$ (curve a),
  $T=10T_{K}$ (curve b), $T=12.5T_{K}$ (curve c) and
  $T=15T_{K}$ (curve d).}
  \label{Fig-GV-SU12-weak}
\end{figure}

\noindent The total differential conductance (\ref{G-SU12-low-D})
is displayed in Fig. \ref{Fig-G-SU12-weak} for $V=0$ (zero bias
differential conductance). The conductance increases when the
temperature is lowered, which is typical to the standard scenario
of
Kondo tunnelling through the tunnel junction.%
\cite{Kaminski-Nazarov-Glazman-00} The nonlinear conductance
(\ref{G-SU12-low-D}) as a function applied bias is shown in Fig.
\ref{Fig-GV-SU12-weak} for several temperatures $T$. The zero bias
peak of the conductance is typical for the ordinary SU(2) Kondo
effect.

%---------------- G(V)-weak ---------------------------

It should be noted that the the conductance (\ref{G-SU12-low-D})
has a factor ${\cal{N}}$, Eq. (\ref{G0-def}),
which is $\frac{3}{4}$ for $N=2$ or $\frac{143}{864}$ for $N=12$.
In other word, as far as the conductance  {\em in the weak coupling regime} 
is concerned, the main
difference between the SU(12) and the SU(2) Kondo tunneling is the
substantial difference of the corresponding Kondo temperatures
(\ref{TK-SU12}).  This similarity no longer holds in the strong coupling
regime as we will now show.

\noindent
\textbf{Conductance in the Strong Coupling Limit:} \\
For $T<T_K$, the mean field slave boson approximation (MFSBA) is
employed to calculate the zero bias tunneling conductance. In the
limit $U\to\infty$, the dot can be either empty or singly
occupied. The dot electron  annihilation  and creation operators
are written as $d_{\lambda}$=$b^{\dag}f_{\lambda}$ and
$d_{\lambda}^{\dagger}$=$f_{\lambda}^{\dag}b$ where the slave
fermion operators $f_{\lambda}$ and the slave boson operator $b$
satisfy the constraint condition,
$$
 Q =
 \sum_{\lambda}
 f_{\lambda}^{\dag}f_{\lambda}+
 b^{\dag}b = 1.
$$
This condition is encoded by including a Lagrange multiplier
$\omega$ in the total action $S$. In the mean field approximation,
we replace the Bose operators $b$ and $b^{\dag}$ by their
expectation values, $b_0=\sqrt{\langle{b^{\dag}b}\rangle}$. At the
mean field level the constraint condition is satisfied only on the
average.

The current operator reads,
\begin{eqnarray}
  I &=&
  \frac{i e b_0}{\hbar}
  \sum_{\lambda}
  t_m
  \Big[
      \psi_{l\lambda}^{\dag}(0)
      f_{\lambda}-
      {\rm{h.c.}}
  \Big].
  \label{I-def}
\end{eqnarray}
It can be derived from the partition function that is  formally
written as,
\begin{equation}
  \label{Z}
  Z(\alpha_q) =
  \int D[f f^\dagger c~c^{\dagger}]
  e^{-\beta S(\alpha_q)}.
\end{equation}
Here $S(\alpha_q)$ is the action (written explicitly below) that
contains a term  $\alpha_q I$ where $\alpha_q$ is a source field,
and integration is carried out over lead ($c,c^\dagger$) and slave
fermion ($f,f^\dagger$)  fields (treated here as Grassman
variables). The action is given explicitly as,
\begin{eqnarray}
  S =
  \int\limits_{-\infty}^{\infty}dt~
  {\cal{L}}(t),
  \label{action-def}
\end{eqnarray}
where ${\cal{L}}={\cal{L}}_l+{\cal{L}}_r+{\cal{L}}_d-{\cal{L}}_t-%
\alpha_qI$,
\begin{eqnarray*}
  {\cal{L}}_{\alpha} &=&
  \sum_{k\lambda}
  c_{\alpha{k}\lambda}^{\dag}
  \Big\{i\hbar\partial_t-\epsilon_{km}\Big\}
  \tau^z
  c_{\alpha{k}\lambda},
  \ \
  \alpha=l,r,
  \\
  {\cal{L}}_d &=&
  \sum_{\lambda}
  f_{\lambda}^{\dag}
  \Big\{i\hbar\partial_t-\epsilon_f\Big\}
  \tau^z
  f_{\lambda},
  \ \ \ \
  \epsilon_f=\epsilon_d+\omega,
  \\
  {\cal{L}}_t &=&
  \frac{b_0}{\sqrt{L_{\rm{cnt}}}}
  \sum_{\alpha{k}\lambda}
  t_m
  \Big\{
      c_{\alpha{k}\lambda}^{\dag}
      \tau^{z}
      f_{\lambda}+
      f_{\lambda}^{\dag}
      \tau^{z}
      c_{\alpha{k}\lambda}
  \Big\}.
\end{eqnarray*}

The  action  in the MFSBA is Gaussian and depends on two real
numbers, the boson field $b_0$ and the chemical potential
(Lagrange multiplier) $\omega$. Carrying out the integration
according to Eq.~(\ref{Z}) yields the partition function,
\begin{eqnarray*}
  \ln Z(\alpha_q) =
  -2
  \sum_{m}
  {\rm{tr}} \ln
  \Big\{
      {\cal{G}}_{fm}^{-1}-
      \frac{e \alpha_q t_m^2 b_0^2}{\hbar}
      \big[
          g_{lm},
          \tau_x
      \big]
  \Big\},
\end{eqnarray*}
where
$$
  {\cal{G}}_{fm}^{-1}=
  g_{f}^{-1}-
  t_m^2 b_0^2
  \Big(g_{lm}+g_{rm}\Big).
$$
Here $g_f$ is the GF (\ref{GF-dot}) of the (non-interacting)
electron in the QD with shifted energy level,
$\epsilon_d\to\epsilon_f=\epsilon_d+\omega$.

\begin{figure}[htb]
%H=6.28, L=14.65
\centering
\includegraphics[width=70 mm,angle=0]
  {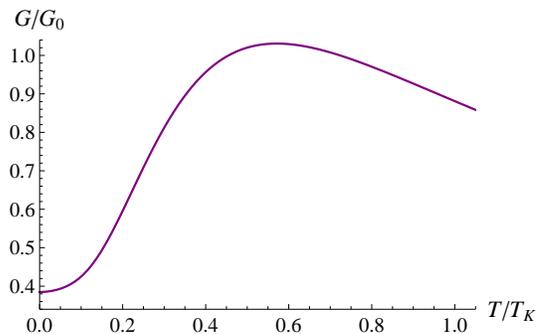}
 \caption{\footnotesize
   (Color online) The zero bias conductance as a function
   of temperature in the strong coupling limit ($T<T_K$).}
\label{Fig-G-strong}
\end{figure}

\noindent The MFSBA is reliable in equilibrium, $V=0$. Therefore
we will consider below the temperature dependence of the zero bias
conductance. In equilibrium, the mean field solutions for $b_0$
and $\omega$ minimize the free energy,
\begin{equation}
  F=
  -2T
  \sum_{m\omega_n}
  {\rm{tr}}~
  \ln
  {\cal{G}}_{fm}^{-1}(i\omega_n)+
  \omega b_0^2,
  \label{Free}
\end{equation}
where the last term is the slave boson kinetic part of the free
energy due to the constraint, and ${\cal{G}}_{fm}^{-1}(i\omega_n)$
is the Matsubara's GF. The mean field equations,
\begin{eqnarray}
  &&
  \frac{N}{\pi}~
  \arctan
  \bigg(
       \frac{b_0^2\Gamma}
            {2\epsilon_f}
  \bigg)
  = 1-b_0^2,
  \nonumber \\ &&
  \frac{N\Gamma}{8\pi}~
  \ln
  \Bigg(
       \frac{\bar{D}^2}
            {\big(\frac{b_0^2\Gamma}{2}\big)^2+\epsilon_f^2}
  \Bigg)
  = \omega,
  ~~~
  \label{MeanFieldEqs}
\end{eqnarray}
are solved for $\omega$ and $b_0$ with the solutions,
$$
  \omega =
  -\epsilon_d+\epsilon_f,
%  T_K
%  \cos
%  \Big(
%      \frac{\pi}{N}
%  \Big)
,
  \ \ \ \ \
  b_0^2 =
  \frac{2 T_K}{ \Gamma}~
  \sin
  \Big(
      \frac{\pi}{N}
  \Big),
$$
where $\epsilon_f=T_K\cos(\pi/N)$, $T_K$ being the Kondo temperature given by Eq. (\ref{TK-SU12})
and $\Gamma$ is given in Eq.~(\ref{width}). The expression for the
linear conductance for $T<T_K$ is now obtained as,
\begin{eqnarray}
  G(T) =
  \frac{N G_0}{8T}
  \int
  \frac{d\epsilon}
       {\displaystyle
        \cosh^2
        \Big(
            \frac{\epsilon}{2T}
        \Big)}~
  \frac{\displaystyle
        \Big(\frac{\pi T_K}{N}\Big)^2}
       {\displaystyle
        \big( \epsilon-\epsilon_f \big)^2+
        \Big(\frac{\pi T_K}{N}\Big)^2},
  \nonumber \\
  \label{G-int}
\end{eqnarray}
with $G_0$ given
by Eq. (\ref{G0-def}). The zero bias conductance as a function of
temperature is shown in Fig.\ref{Fig-G-strong}. It is seen that
the conductance has a peak at  $T\approx0.57T_K$  due to
the constraint imposed by the Friedel sum
rule.\cite{Hewson-book,Rajan83} In addition to the different Kondo
temperatures (\ref{TK-SU12}) for the SU(12) and SU(2) Kondo effects,
this behavior indicates a remarkable distinction from the standard SU(2) Kondo
tunneling.\cite{KEQD} In the latter case, the conductance is monotonically
increasing towards the unitary limit as $T \to 0$. It should be noted that
we define $G_0$ as  $e^2/(6h)$ {\it per spin projection} [see eq.(\ref{G0-def})],
so that the unitary limit corresponds here to $6G_0=12e^2/h$.
A close inspection
shows that this limit is not perfectly reached. The reason is that
while the DOS has a peak that is shifted from the Fermi level by  $T_K$,
 the peak of the "thermal" function
$\cosh^{-2}(\epsilon/2T)$ sits right
at the Fermi level. As a result, the peak
of the conductance occurs at finite temperature, and its value is slightly lowered by
the "thermal" function.

%---------------- conduct-strong -------------------------

\section{Magnetic Susceptibility}
  \label{sec-magn}
While in bulk metals, the Kondo effect manifests itself through
measurements of electrical resistivity {\it and} magnetic
susceptibility, in quantum dots it manifests itself mainly through
the properties of the conductance. Designing experiments aiming at
studying magnetic response of quantum dot in the Kondo regime is
rather difficult because they require an STM technique in which
the tip is close to the magnetic impurity. Appropriate STM
techniques have already been worked out for impurities composed of
added magnetic atoms on metallic surface.\cite{Crommie} We are
unaware of their applications in quantum dots. The discussion
below is therefore motivated by our hope that measurement of
magnetic response of a single magnetic impurity in quantum dot
will eventually materialized.

In the CNT-CNT(QD)-CNT junction the magnetic response is encoded
by the static impurity magnetic susceptibility $\chi$ of the
CNT(QD) (defined explicitly below). Unlike the discussion
pertaining to the conductance, there is no source-drain bias
present here, and the leads just serve as a source of electron gas
that acts to screen the impurity. The distinction between the
present structure and that of Kondo effect in bulk
CNT\cite{Cobden} is that here the impurity is composed of a
trapped electron with a 12-fold degenerate ground state.

The Zeeman splitting $\Delta_{m\sigma}$ of electron energy levels
in a carbon nanotube subject to an external magnetic field
${\bf{B}}$ depends on whether the magnetic field is parallel or
perpendicular to the CNT axis (see Appendix \ref{append-Zeeman}
for details). Explicitly,
\begin{eqnarray}
  \Delta_{m\sigma} &=&
  -g_{\rm{orb}} m \mu_B B_{\parallel}
  -g_{\rm{spin}} \sigma \mu_B B,
  \label{Zeeman-split-def}
\end{eqnarray}
where $B=|{\bf{B}}|$, $B_{\parallel}$ is the component of the
magnetic field parallel to the CNT axis, $\mu_B$ is the Bohr
magneton, $g_{\rm{orb}}$ and $g_{\rm{spin}}$ are orbital and spin
$g$-factors,
\begin{equation}
  g_{\rm{spin}} \approx 2,
  \ \ \ \ \
  g_{\rm{orb}} =
  \frac{m_evr_0}{\pi\hbar}~
  \frac{\Delta_0}{\epsilon_F},
  \label{gorb-gspin-def}
\end{equation}
where $m_e$ is the mass of free electron.

The Zeeman splitting (\ref{Zeeman-split-def}) results in an
anisotropy of the magnetic susceptibility: In other words, $\chi$
is a tensor, which in the principal frame of the CNT  has parallel
and perpendicular components, $\chi_{\rm{imp}}^{\parallel}$ and
$\chi_{\rm{imp}}^{\perp}$,
responding to the magnetic field parallel or perpendicular to the
CNT axis. This anisotropy is absent in the ordinary SU(2) Kondo
effect, and is one of the hallmarks of a higher symmetry such as
SU(12) discussed here that involves orbital symmetry.

The impurity magnetization is defined through the
relation\cite{Hewson-book}
\begin{eqnarray}
  &&
  {\bf{M}}_{\rm{imp}} =
  g_{\rm{spin}} \mu_B
  \bigg\{
       \Big\langle
           {\bf{S}}+
           \sum_{\alpha}
           \boldsymbol\Sigma_{\alpha}
       \Big\rangle-
       \Big\langle
           \sum_{\alpha}
           \boldsymbol\Sigma_{\alpha}
       \Big\rangle_{0}
  \bigg\}+
  \nonumber \\ && ~~ +
  g_{\rm{orb}} \mu_B
  {\bf{e}}_{x}
  \bigg\{
      \Big\langle
        L^x+
        \sum_{\alpha}
        \Lambda_{\alpha}^x
      \Big\rangle-
      \Big\langle
        \sum_{\alpha}
        \Lambda_{\alpha}^x
      \Big\rangle_0
  \bigg\},
  \label{magn-def}
\end{eqnarray}
where ${\bf{S}}$ and $\boldsymbol\Sigma_{\alpha}$ [$\alpha=l,r$]
are respectively the spin operators of the dot and  the lead
electrons,
\begin{eqnarray}
  &&
  {\bf{S}} =
  \sum_{\lambda\lambda'}
  {\bf{s}}_{\sigma\sigma'}
  \delta_{mm'}
  \delta_{\xi\xi'}
  X^{\lambda\lambda'},
  \nonumber \\
  &&
  \boldsymbol\Sigma_{\alpha} =
  \sum_{k\lambda\lambda'}
  c_{\alpha k \lambda}^{\dag}
  {\bf{s}}_{\sigma\sigma'}
  \delta_{mm'}
  \delta_{\xi\xi'}
  c_{\alpha k \lambda'},
  \label{spin-dot-lead-def}
\end{eqnarray}
while $L^x$ and $\Lambda_{\alpha}^{x}$ are respectively the
operators of the $x$-component of the orbital moment of the dot or
the lead,
\begin{eqnarray}
  L^x =
  \sum_{\lambda}
  m
  X^{\lambda\lambda},
  \ \ \
  \Lambda_{\alpha}^{x} =
  \sum_{k\lambda}
  m
  c_{\alpha k \lambda}^{\dag}
  c_{\alpha k \lambda}.
  \label{orb-dot-lead-def}
\end{eqnarray}
In Eq.~(\ref{magn-def}), $\langle\ldots\rangle$ indicates thermal
averaging with respect to the full Hamiltonian $\tilde{H}=H_0+H_K$
[equations (\ref{H-Anderson-def}) and (\ref{HK-SU12-def})],
whereas $\langle\ldots\rangle_{0}$ indicates thermal averaging
respect to $H_0$. It is reasonably
assumed that electrons in the dot and the leads have the same $g$-factors.\\

\noindent
{\bf Susceptibility in the weak coupling regime}\\
Using a similar analysis as for the conductance, we derive an
expression for the zero-field magnetic susceptibility to second
order in $j$,
\begin{eqnarray}
  \chi_{\rm{imp}}^{\parallel} &=&
  \bigg(
       \frac{g_{\rm{spin}}^2}{4}+
       \frac{2g_{\rm{orb}}^2}{3}
  \bigg)~
  \chi(T),
  \label{chi-parallel}
  \\
  \chi_{\rm{imp}}^{\perp} &=&
  \frac{g_{\rm{spin}}^2}{4}~
  \chi(T),
  \label{chi-perp}
\end{eqnarray}
where, to second order in $j$,
\begin{eqnarray}
  &&
  \chi(T) =
  \frac{\chi_0T_K}{T}~
  \bigg\{
       1-
       j-
       \frac{Nj^2}{2}~
       \ln
       \bigg(
            \frac{D}{T}
       \bigg)
  \bigg\},
  \label{suscept-sec-order}
  \\
  &&
  \chi_0 =
  \frac{\mu_B^2}{T_K}.
  \label{chi0-def}
\end{eqnarray}
The second term on the RHS of Eq.~(\ref{chi-parallel}) reflects
the orbital degeneracy, and is absent in the SU(2) Kondo effect.
This anisotropy of the magnetic response is one of our main
results, as it constitutes an observable that is a hallmark of the
SU(12) symmetry of the pertinent Kondo effect. It is compactly
encoded by the temperature independent ratio,
\begin{equation}
  \label{ratio}
  \frac{\chi_{\rm{imp}}^{\parallel}}
       {\chi_{\rm{imp}}^{\perp}}=
  1+
  \frac{8}{3}~
  \frac{g_{\rm{orb}}^2}{g_{\rm{spin}}^2}.
\end{equation}
As we shall see below, this relation holds also in the strong coupling
regime $T<T_K$.
It is then suspected that this result is ``universal" in the sense that
it holds for the crossover region $T \approx T_K$ as well.
In Appendix \ref{append-fluct-dissip} it is indeed shown that this ratio
can be derived quite generally  (in this model) by
using the fluctuation-dissipation formula for the susceptibility (which relates the susceptibility
to the spin correlations).

At high temperatures, the logarithmic term causes a reduction of
the effective magnetic moment as compared with that for a free
spin. With decreasing temperature, the second order perturbation
theory becomes inadequate. In order to derive an expression for
$\chi_{\rm{imp}}$ in the leading logarithmic approximation, we use
the RG equations (\ref{scale-eq-SU12}). The condition imposing the
invariance of the susceptibility under the poor man's scaling
transformation is,
\begin{eqnarray*}
  \frac{\chi_0T_K}{T}~
  \frac{\partial}{\partial\ln(D)}
  \bigg\{
       1-
       j-
       \frac{Nj^2}{2}
       \ln
       \bigg(
            \frac{D}{T}
       \bigg)
  \bigg\}
  &=& 0.
\end{eqnarray*}
Within the accuracy of this equation, when differentiating the
third term, we should neglect any implicit dependence on $D$
through the coupling $j$. The renormalization procedure should
proceed until the bandwith $D$ is reduced to the temperature $T$.
At this point, the second order of the perturbation theory
vanishes and the susceptibility takes the form,
\begin{eqnarray}
  \chi(T) &=&
  \frac{\chi_0T_K}{T}~
  \bigg\{
       1-
       \frac{2}{N\ln(T/T_K)}
  \bigg\}.
  \label{suscept-RG}
\end{eqnarray}

\begin{figure}[htb]
 \centering
 \includegraphics[width=65 mm,angle=0]
   {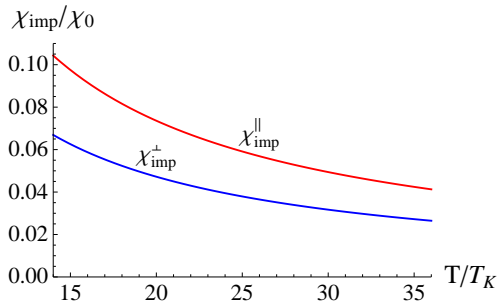}
 \caption{\footnotesize
  (Color online) Impurity susceptibility
  $\chi^{\parallel}_{\rm{imp}}$ [red curve] and
  $\chi^{\perp}_{\rm{imp}}$ [blue curve],
  as a function of temperature in the weak coupling regime
  [equations (\ref{chi-parallel}), (\ref{chi-perp}) and (\ref{suscept-RG})].}
  \label{Fig-suscept}
\end{figure}

The impurity susceptibility in the weak coupling regime, equations
(\ref{chi-parallel}), (\ref{chi-perp}) and (\ref{suscept-RG}), is
shown in Fig.\ref{Fig-suscept}.

%---------------- susceptibility weak ---------------------------

\noindent {\bf{Susceptibility in the strong coupling regime}}: For
$T<T_K$, the magnetic susceptibility can be calculated in the
framework of the MFSBA. For this purpose, we take into account the
dependence of the right hand side of Eq.~(\ref{Free}) for the free
energy on the external magnetic field ${\bf B}$. Because the
susceptibility tensor is diagonal, we may write
$\chi_{\rm{imp}}^{i}=-[\partial^2 F({\bf B})/\partial B_i^2]_{{\bf B}=0}$, where
$i=\parallel,\perp$. Thereby  we get the zero field susceptibility
$\chi_{\rm{imp}}^{\parallel}$ or $\chi_{\rm{imp}}^{\perp}$. Explicitly,
for magnetic field parallel or perpendicular to the CNT axis, the susceptibility is
given by equation (\ref{chi-parallel}) or (\ref{chi-perp}), with
$\chi(T)$ given by
\begin{eqnarray}
  \chi(T) =
  \frac{\chi_0}{4T}~
  \int
  \frac{d\epsilon}
       {\displaystyle
        \cosh^2\left(\frac{\epsilon}{2T}\right)}~
  \frac{\displaystyle
        \Big(\frac{T_K}{N}\Big)^2}
       {\displaystyle
        \big(\epsilon+T_K\big)^2+
        \Big(\frac{\pi T_K}{N}\Big)^2}.
  \nonumber \\
  \label{chi-strong}
\end{eqnarray}

%---------------- susceptibility-strong ---------------------------
\begin{figure}[htb]
 \centering
 \includegraphics[width=65 mm,angle=0]
   {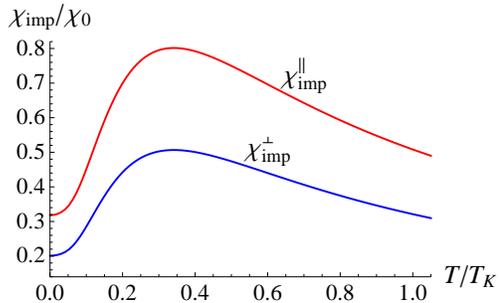}
 \caption{\footnotesize
  (Color online) Impurity susceptibility $\chi^{\parallel}_{\rm{imp}}$
  [red curve] and $\chi^{\perp}_{\rm{imp}}$ [blue curve]
  as a function of temperature in the strong coupling regime
  [equations (\ref{chi-parallel}), (\ref{chi-perp}) and (\ref{chi-strong})].}
  \label{Fig-suscept-strong}
\end{figure}

The magnetic susceptibilities in the strong coupling regime are
shown in Figure \ref{Fig-suscept-strong}.  They display a
peak at finite-temperature, commensurate with the constraint
imposed by the Friedel sum rule.\cite{Hewson-book,Rajan83,Andrei}

%%%%%%%%%%%%%%%%%%%%%%%%%%%%%%%%%%%%%%%%%%%%%%%%%%%%%%%%%%%

\section{Conclusions}
  \label{sec-concl}
Whereas the theoretical framework of the Coqblin-Schrieffer 
model is intensively studied, the present work focuses on one of its special facet 
that is less explored, namely,  its possible realization 
in a transport device with a Dirac spectrum and peculiar DOS.
We substantiate the possibility of tuning a metallic CNT into a
tunnel junction consisting of two CNT metallic leads and a
CNT(QD). The spin, isospin (valley) and orbital degeneracy of the
CNT(QD) energy spectrum gives rise to the Kondo effect with SU(12)
dynamical symmetry. The high symmetry of the CNT(QD) leads to an
enhanced Kondo temperature. The conductance through the junction
is evaluated using Keldysh technique. Renormalization group
analysis is performed in the weak coupling regime ($T \gg T_K$)
while the MFSBA is used at the strong coupling regime $T<T_K$.
  In the weak coupling regime, the behavior of $G(T)$ as
function of temperature for the SU(12) Kondo effect is
qualitatively the same as that for the ordinary SU(2) Kondo
effect, and the main difference is that $T_K$[SU(12)]$ \gg
T_K$[SU(2)].  In the strong coupling regime the situation is different.
Due the constraints imposed by the Friedel sum rule, the
conductance has a peak at finite temperature that becomes sharper
the higher is $N$. This distinction of the conductance between SU(2) and SU(12)
Kondo effect in quantum dot should be experimentally observable.

The magnetic response exposes yet another
remarkable distinction between the SU(12) and the SU(2) Kondo
effects. For the SU(12) Kondo effect, the response is anisotropic
and the susceptibility is a tensor. It has two components,
$\chi_{\rm{imp}}^{\parallel}$ and $\chi_{\rm{imp}}^{\perp}$ according
to whether the magnetic field is along the CNT axis or perpendicular to it.
Moreover, the ratio $\chi_{\rm{imp}}^{\parallel}/\chi_{\rm{imp}}^{\perp}=
1+8g_{\mathrm{orb}}^2/(3 g_{\mathrm{spin}}^2)$ depends only on the
orbital and spin $g$ factors. This result is demonstrated in the
weak coupling regime based on RG calculations  and in the strong
coupling regime based on the MFSBA. A proof that this result is
true in every order of perturbation theory is derived in
Appendix \ref{append-fluct-dissip}
employing the fluctuation-dissipation theorem.
 An experimental search for such anisotropy would
constitute a confirmation of this unusual Kondo effect,  but
as was pointed out earlier, observing magnetic response of a
single impurity is quite difficult. 

The Kondo physics in systems with Dirac spectrum proves to be
rather rich. While the Kondo effect in bulk graphene reveals
peculiar equilibrium properties such as the existence of two
distinct classes of Kondo quantum critical points\cite{neto},
analysis of non-equilibrium transport  in correlated CNT(left
lead)-CNT(QD)-CNT(right lead) junction reveals another facet,
namely,  Kondo tunneling with an SU(12) dynamical symmetry.
\\
\ \\
{\bf Acknowledgment:} We would like to thank Natan Andrei for
stimulating discussions. Correspondence with George Martins is
highly appreciated. This research was supported by the Israeli
Science Foundation for supporting our research under grants 1173/08
and 400/2012.

%%%%%%%%%%%%%%%%%%%%%%%%%%%%%%%%%%%%%%%%%%%%%%%%%%%%%%%%%%%

\appendix

%%%%%%%%%%%%%%%%%%%%%%%%%%%%%%%%%%%%%%%%%%%%%%%%%%%%%%%%%%%

%%%%%%%%%%%%%%%%%%%%%%%%%%%%%%%%%%%%%%%%%%%%%%%%%%%%%%%%%%%
\section{Wave Functions of CNT Quantum Dot}
  \label{append-dot}

For electrons in CNT(QD), the single electron wave functions and
the corresponding energy spectrum are derived from the Dirac
equation (\ref{Dirac-eq-dot}). The solution $\Phi_{mn}(x,\phi)$ of
the Dirac equation is written as,
\begin{eqnarray}
  \Phi_{mn}(x,\phi) =
  \left\{
  \begin{array}{lcr}
  \Phi_{mn}^{(1)}(x)~e^{im\phi} & {\rm{if}} & |x|<h,
  \\
  \Phi_{mn}^{(2)}(x)~e^{im\phi} & {\rm{if}} & x>h,
  \\
  \Phi_{mn}^{(3)}(x)~e^{im\phi} & {\rm{if}} & x<-h,
  \end{array}
  \right.
  \label{phiphi-append}
\end{eqnarray}
where $m=0,\pm1,\pm2,\ldots$~ is a magnetic quantum number,
$n=0,1,2,\ldots$ is a radial quantum number.

The function $\Phi_{mn}^{(1)}(x)$ is given by,
\begin{eqnarray}
  \Phi_{mn}^{(1)}(x)
  &=&
  \frac{A_{mn}}
       {\sqrt{\epsilon}}~
  \bigg\{
       \chi^{(1)}_{\up}
       \sqrt{\epsilon+M_m}
       \cos
       \Big(
           k_x x+
           \frac{n\pi}{2}
       \Big)+
  \nonumber \\ && +
       i\chi^{(1)}_{\down}
       \sqrt{\epsilon-M_m}
       \sin
       \Big(
           k_x x+
           \frac{n\pi}{2}
       \Big)
  \bigg\},
  \label{phi-in-append}
\end{eqnarray}
where $M_m$ and $M_0$ are defined through the relations
$$
  \hbar v k_x =
  \sqrt{\epsilon^2-M_m^2},
  \ \ \ \ \
  M_m =
  \sqrt{M_0^2+m^2\Delta_0^2}.
$$
The expressions for the spinors $\chi^{(1)}_{\sigma}$ and
$\chi^{(2)}_{\sigma}$ (to be used later) are given in equation
(\ref{chi-1-2-up-def}).

The function $\Phi_{m}^{(1)}(x)$ has the following symmetry,
\begin{eqnarray*}
  &&
  \hat{M}_m
  \Phi_{m}^{(1)}(-x) =
  (-1)^{n}
  M_m
  \Phi_{m}^{(1)}(x).
\end{eqnarray*}

Similarly, the function $\Phi_{m}^{(2)}(x)$ (for $x>h$) reads,
\begin{eqnarray}
  \Phi_{mn}^{(2)}(x) =
  \frac{B_{m}e^{-\kappa(x-h)}}{\sqrt{2N_m}}
  \times ~~~~~~~~~~ ~~~~~~~~~~
  \nonumber \\ \times
  \bigg\{
      \chi^{(2)}_{\up}
       \sqrt{N_m+\epsilon}+
       i\chi^{(2)}_{\down}
       \sqrt{N_m-\epsilon}
  \bigg\},
  \label{Phi-out-2-append}
\end{eqnarray}
where $N_m$ and $N_0$ are defined through
$$
  \hbar v \kappa =
  \sqrt{N_m^2-\epsilon^2},
  \ \ \ \ \
  N_m =
  \sqrt{N_0^2+m^2\Delta_0^2}.
$$
Finally, the function $\Phi_{mn}^{(3)}(x)$ (for $x<-h$) is
\begin{eqnarray*}
  &&
  \Phi_{mn}^{(3)}(x)
  =
  \frac{(-1)^{n}}{N_m}~
 \hat{N}_m
  \Phi_{mn}^{(2)}(-x).
  \nonumber
\end{eqnarray*}

Applying the continuity condition for $\Phi_{mn}(x,\phi)$,
Eq.~(\ref{phiphi-append}) at the points $x=\pm{h}$, we obtain the
set of equations,
\begin{subequations}
\begin{eqnarray}
  A_{mn}
  \sqrt{\epsilon+M_m}~
  \cos
  \Big(
      k_x h+
      \frac{n\pi}{2}
  \Big)
  =
  B_{mn}
  {\cal{F}}_{\up}(\epsilon),
  \nonumber \\
  \label{eq1-cont-append}
  \\
  A_{mn}
  \sqrt{\epsilon-M_m}~
  \sin
  \Big(
      k_x h+
      \frac{n\pi}{2}
  \Big)
  =
  B_{mn}
  {\cal{F}}_{\down}(\epsilon),
  \nonumber \\
  \label{eq2-cont-append}
\end{eqnarray}
  \label{set-eqs-cont-append}
\end{subequations}
where ${\cal{F}}_{\sigma}(\epsilon)$ is given by
Eq.(\ref{cal-F-def}).

The set of equations (\ref{set-eqs-cont-append}) has nontrivial
solutions when its determinant vanishes. This condition gives us
equation (\ref{Eq-for-spectrum}) for the energy levels in the
quantum dot.

%%%%%%%%%%%%%%%%%%%%%%%%%%%%%%
\section{Magnetization of the Tunnel Junction}
  \label{append-Zeeman}

It order to describe electronic properties of a carbon nanotube in
an external magnetic field ${\bf{B}}$, we should add to the CNT
Hamiltonian the term $H_{B}$ describing spin-Zeeman splitting,
\begin{equation}
  H_{\rm{spin}} =
  -g_{\rm{spin}}
  \mu_B
  \big(
      {\bf{s}}
      \cdot
      {\bf{B}}
  \big),
  \label{H-spin-append-def}
\end{equation}
and replace the wave vector ${\bf{k}}$ by the operator
${\bf{k}}'$,\cite{Ando-93-1,Ando3}
\begin{eqnarray*}
  {\bf{k}} \to {\bf{k}}' =
  -i\nabla-
  \frac{e}{\hbar c}~
  {\bf{A}}.
\end{eqnarray*}
Here ${\bf{s}}$ is a vector of the spin operators, ${\bf{A}}$ is a
vector potential, ${\bf{B}}=\nabla\times{\bf{A}}$. Then the motion
of electron in a CNT with the wave vector close to the ${\bf{K}}$
point of the first Brillouin zone can be described by the
Hamiltonian,
\begin{eqnarray}
  H =
  \hbar v
  \Big(
      {\bf{k}}-
      \frac{e}{\hbar c}~
      {\bf{A}}
  \Big)
  \cdot
  \boldsymbol\tau+
  \Delta_g\tau_z+
  H_{\rm{spin}}.
  \label{H-append-def}
\end{eqnarray}
Here we use use the cylindrical system of coordinates where
${\bf{k}}=(k_x,k_y)$ with $k_x=-i\partial_x$ and
$k_y=-\frac{i}{r_0}\partial_{\phi}$. The Hamiltonian for the
motion of electron with the wave vector near ${\bf{K}}'$ point can
be obtained from equation (\ref{H-append-def}) just by replacing
${k_y}\to{-k_y}$.

In what follows, we will calculate Zeeman splitting for the
magnetic field parallel and perpendicular to the CNT axis.

\noindent {\bf{Magnetic field parallel to the CNT axis}}: When the
magnetic field is parallel to the CNT axis,
${\bf{B}}_{\|}=B{\bf{e}}_x$, the vector potential can be written
as,
\begin{eqnarray}
  {\bf{A}}_{\|} &=&
  \frac{B r}{2}~
  {\bf{e}}_{\phi}.
  \label{A-par-def}
\end{eqnarray}
The eigenfunction of the Hamiltonian (\ref{H-append-def}) is,
\begin{eqnarray}
  \big|\Psi_{skm\sigma}^{\parallel}(\varphi)\big\rangle &=&
  \big|\chi_{\sigma}\big\rangle
  \otimes
  \big|\psi_{skm}(\varphi)\big\rangle,
  \label{Psi-par-append-def}
\end{eqnarray}
where $\varphi=\pi B r_0^2$ is the magnetic flux through the cross
section of the CNT. Here $|\chi_{\sigma}\rangle$ is a spin wave
function of electron with spin parallel or anti-parallel to the
magnetic field,
\begin{eqnarray}
  &&
  \big|\chi_{\up}\big\rangle =
  \left(
  \begin{array}{c}
  1 \\ 0
  \end{array}
  \right),
  \ \ \ \ \
  \big|\chi_{\down}\big\rangle =
  \left(
  \begin{array}{c}
  0 \\ 1
  \end{array}
  \right).
  \label{chi-z}
\end{eqnarray}
$|\psi_{skm}(\varphi)\rangle$ is the spatial wave function of
electron in the conduction ($s=+1$) or valence ($s=-1$) band with
orbital quantum number $m$ ($m=0,\pm1$), and wave number $k$,
\begin{eqnarray}
  &&
  \big|\psi_{s k m}(\varphi)\big\rangle =
  \frac{e^{i k x+i m \phi}}
        {\sqrt{4 \pi L}}~
  \left(
  \begin{array}{c}
  s b_{km}(\varphi) \\ 1
  \end{array}
  \right),
  \label{psi-orb-par}
  \\
  &&
  b_{km}(\varphi) =
  \frac{\kappa_m(\varphi)-ik}{\sqrt{\kappa_m^2(\varphi)+k^2}},
  \ \ \ \ \
  \kappa_m(\varphi)=\frac{m-\varphi}{r_0}.
  \nonumber
\end{eqnarray}

The corresponding energy is,
\begin{eqnarray}
  \tilde\veps_{sk\lambda} &=&
  s
  \sqrt{(\hbar v k)^2+(m-\varphi)^2\Delta_0^2+\Delta_g^2}-
  \nonumber \\ && -
  2 \sigma \mu_B B.
  \label{energy-smk}
\end{eqnarray}
For weak magnetic fields ($\varphi\ll1$),
$\varepsilon_{sk\lambda}$ can be expanded to linear with $B$
correction,
\begin{eqnarray*}
  &&
  \tilde\veps_{sk\lambda} =
  \veps_{skm}-
  \frac{\Delta_0^2 m \varphi}{\varepsilon_{skm}}-
  2 \sigma \mu_B B+
  O(\varphi^2),
  \\
  &&
  \veps_{skm} =
  s
  \sqrt{(\hbar v k)^2+m^2\Delta_0^2+\Delta_g^2}.
\end{eqnarray*}
Then for $\varepsilon_{skm}$ close to the Fermi level, we get
equation (\ref{Zeeman-split-def}).

Then the magnetization (\ref{magn-def}) in linear with $B$
approximation is,
\begin{eqnarray}
  M_{\rm{imp}}^x &=&
  g_{\rm{spin}}^2 \mu_B^2 B
  \bigg\{
       \Big\langle
           \big(\tilde\Sigma^x\big)^2
       \Big\rangle-
       \Big\langle
           \sum_{\alpha}
           \big(\Sigma_{\alpha}^x\big)^2
       \Big\rangle_{0}
  \bigg\}+
  \nonumber \\ &+&
  g_{\rm{orb}}^2 \mu_B^2 B
  \bigg\{
      \Big\langle
          \big(\tilde\Lambda^{x}\big)^2
      \Big\rangle-
      \Big\langle
          \sum_{\alpha}
          \big(\Lambda_{\alpha}^x\big)^2
      \Big\rangle_{0}
  \bigg\},
  \nonumber \\
  \label{magnetization-par-lin}
\end{eqnarray}
where
$\tilde{\boldsymbol\Sigma}=(\tilde\Sigma^x,\tilde\Sigma^y,\tilde\Sigma^z)$
is the total spin of the tunnel junction,
\begin{eqnarray}
  \tilde{\boldsymbol\Sigma} &=&
  {\bf{S}}+
  \sum_{\alpha}
  \boldsymbol\Sigma_{\alpha},
  \label{spin-total-def}
\end{eqnarray}
$\tilde\Lambda^x$ is the orbital momentum of the total system,
\begin{eqnarray}
  \tilde\Lambda^x &=&
  L^x+
  \sum_{\alpha}
  \Lambda_{\alpha}^x.
\end{eqnarray}

\noindent {\bf{Magnetic field perpendicular to the CNT axis}}: Let
us consider now the magnetic field perpendicular to the CNT axis.
For definiteness, we take
${\bf{B}}_{\perp}=B[{\bf{e}}_r\cos\phi-{\bf{e}}_{\phi}\sin\phi]$,
so that ${\bf{A}}_{\perp}=Br\sin\phi~{\bf{e}}_x$. Then the
Hamiltonian (\ref{H-append-def}) takes the form,
\begin{eqnarray}
  &&
  H =
  H_0+H_{\rm{spin}}+H_{\rm{orb}},
  \label{H-append-perp}
  \\
  &&
  H_0 =
  \hbar v
  \boldsymbol\tau
  \cdot
  {\bf{k}}+
  \Delta_g
  \tau_z,
  \nonumber
\end{eqnarray}
where $H_{\rm{spin}}$ is given by equation
(\ref{H-spin-append-def}),
\begin{eqnarray}
  H_{\rm{orb}} &=&
  -\frac{\Delta_0 r_0^2}{l_B^2}~
  \sin\phi~
  \tau_x,
  \label{HO-perp}
\end{eqnarray}
$l_B$ is the magnetic length given by
\begin{eqnarray}
  l_B &=&
  \sqrt{\frac{c\hbar}{B e}}.
  \label{Lb-def}
\end{eqnarray}
When $l_B\gg{r_0}$, the field can be regarded as a small
perturbation.

The eigenfunction of the Hamiltonian $H_0+H_{\rm{spin}}$ are
$|\chi_{\sigma}\rangle\otimes|\psi_{skm}\rangle$, where
$|\chi_{\sigma}\rangle$ describes the quantum state with the spin
parallel or anti-parallel to the magnetic field ${\bf{B}}$,
$|\psi_{skm}\rangle$ is the spatial wave function of electron with
wave number $k$, orbital number $m$ in the conduction  or valence
band, $s=\pm1$.

In order to estimate the contribution of $H_{\rm{orb}}$, we note
that the nontrivial matrix elements of $H_{\rm{orb}}$ are,
$$
  \langle\psi_{skm}|
    H_{\rm{orb}}
  |\psi_{skm+1}\rangle,
  \ \ \ \ \
  \langle\psi_{skm+1}|
    H_{\rm{orb}}
  |\psi_{skm}\rangle,
$$
i.e., $H_{\rm{orb}}$ change the orbital quantum number by $\pm1$
keeping the other quantum numbers (wave number, band index, spin,
...) unchanged. The quantum transitions from the state
$|\psi_{skm}\rangle$ to the state $|\psi_{skm+1}\rangle$ costs the
energy $\veps_{skm+1}-\veps_{skm}\sim\Delta_0$. As a result, for
low magnetic fields [$l_B\gg{r_0}$], corrections of $H_{\rm{orb}}$
to the energy spectrum is of order $l_B^{-4}\sim{B^2}$ and the
energy dispersion in linear with $B$ approximation is given by
equation (\ref{Zeeman-split-def}).

The magnetization (\ref{magn-def}) in linear with $B$
approximation is,
\begin{eqnarray}
  M_{\rm{imp}}^{\perp} =
  g_{\rm{spin}}^2 \mu_B^2 B
  \bigg\{
       \Big\langle
           \big(\tilde\Sigma_{\alpha}^{y}\big)^2
       \bigg\rangle-
       \Big\langle
           \sum_{\alpha}
           \big(\Sigma_{\alpha}^{y}\big)^2
       \Big\rangle_{0}
  \bigg\},
  \nonumber \\
  \label{magnetization-perp-lin}
\end{eqnarray}
where we take the $y$-component of the spin operators for
definiteness, $\tilde\Sigma^y$ is given by equation
(\ref{spin-total-def}).

\section{Magnetic Susceptibility of CNT QD:
       Fluctuation-Dissipation Theorem}
  \label{append-fluct-dissip}

In this section we derive the universal relation, (\ref{ratio}) using
the fluctuation dissipation theorem.

\subsection{Magnetic Susceptibility}
  \label{subsec-susceptibility}

According to the fluctuation-dissipative theorem, the tensor of the magnetic
susceptibility of the quantum dot is defined as,
\begin{eqnarray}
  &&
  \chi_{ij} =
  -\frac{\partial^2F}
        {\partial B_i \partial B_j}=
  \frac{1}{T}~
  \bigg\{
       \big\langle
           m_i
           m_j
       \big\rangle  -\big\langle
           m_i
       \big\rangle~
       \big\langle
           m_j
       \big\rangle \nonumber \\
  && -
       \Big\langle
           m_i^{(0)}
           m_j^{(0)}
       \Big\rangle_0+
       \Big\langle
           m_i^{(0)}
       \Big\rangle_0~
       \Big\langle
           m_j^{(0)}
       \Big\rangle_0
  \bigg\}.
  \label{susceptibility-fluct-diss-def}
\end{eqnarray}
Here $\langle\ldots\rangle$ denotes the thermal average with respect
to the Hamiltonian of interacting quantum dot and leads,  $\langle\ldots\rangle_0$
is the average with respect to the Hamiltonian of the isolated leads.
$i,j=x,y,z$ are Cartesian indices, ${\bf{m}}=(m_x,m_y,m_z)$
is magnetic momentum of the quantum dot and the lead,
${\bf{m}}^{(0)}=\left(m_x^{(0)},m_y^{(0)},m_z^{(0)}\right)$ is magnetic moment of
isolated leads,
\begin{eqnarray}
  &&
  {\bf{m}} =
  g_{\rm{spin}} \mu_B
  \bigg\{
       {\bf{S}}+
       \sum_{\alpha}
       \boldsymbol\Sigma_{\alpha}
  \bigg\} \nonumber \\
  && +
  g_{\rm{orb}} \mu_B
  {\bf{e}}_{x}
  \bigg\{
       L^x+
       \sum_{\alpha}
       \Lambda_{\alpha}^x
  \bigg\},
  \label{magn-operator-def}
  \\
  &&
  {\bf{m}}^{(0)} =
  g_{\rm{spin}} \mu_B
  \sum_{\alpha}
  \boldsymbol\Sigma_{\alpha}+
  g_{\rm{orb}} \mu_B
  {\bf{e}}_{x}
  \sum_{\alpha}
  \Lambda_{\alpha}^x,
  \label{magn0-operator-def}
\end{eqnarray}
where the spin operators of the dot and  the lead
electrons
[${\bf{S}}$ and $\boldsymbol\Sigma_{\alpha}$, $\alpha=l,r$]
are given by Eq. (\ref{spin-dot-lead-def}),
while the operators of the $x$-component of the orbital
moment of the dot and the lead [$L^x$ and $\Lambda_{\alpha}^{x}$,
$\alpha=l,r$] are given by Eq. (\ref{orb-dot-lead-def}).
The Hubbard operator $X^{\lambda \lambda'}=|\lambda \ra \la \lambda'|$
is defined after Eq.~(\ref{HK-SU12-def1}).
It is reasonably assumed that electrons in the dot and the leads
have the same $g$-factors.

We choose the set of coordinates in such a way that the $x$ axis
is parallel to the CNT axis, whereas the $y$ and $z$ axes are
perpendicular. In this set of coordinates, the tensor of the susceptibility
is diagonal,
\begin{eqnarray*}
  \hat\chi &=&
  \left(
  \begin{array}{ccc}
  \chi_{\parallel} & 0 & 0 \\
  0 & \chi_{\perp} & 0 \\
  0 & 0 & \chi_{\perp}
  \end{array}
  \right).
\end{eqnarray*}
{\it We will prove that the zero-field susceptibilities satisfy the ratio,}
\begin{eqnarray}
  \frac{\chi_{\parallel}}
       {\chi_{\perp}}
  &=&
  1+
  \frac{8}{3}~
  \frac{g_{\rm{orb}}^2}{g_{\rm{spin}}^2}.
  \label{ratio-suscept-proof}
\end{eqnarray}

For this purpose, we note the following:
The Kondo Hamiltonian (\ref{HK-SU12-def}) describes the co-tunneling
process such that an electron with the quantum number $\lambda$
(the spin $\sigma$, the orbital quantum number $m$ and the valley
number $\xi$) exits from the dot to the lead and another electron
with the quantum number $\lambda'$ (the spin $\sigma'$, the orbital
quantum number $m'$ and the valley number $\xi'$) enters the quantum
dot from the lead. That mean that the total spin and the total
orbital momentum of the lead and the quantum dot are the good
quantum numbers.

\noindent
{\bf{Proof}}: \underline{Let us consider first $\chi_{\perp}$},
\begin{eqnarray}
&&  \chi_{\perp} \mbox{=}
  \frac{g_{\rm{spin}}^2\mu_B^2}{T}
  \bigg\{
       \Big\langle
            S^zS^z\mbox{+}
            2\sum_{\alpha}
            S^z\Sigma_{\alpha}^z \nonumber \\
            &&+
            \sum_{\alpha\alpha'}
            \Sigma_{\alpha}^z
            \Sigma_{\alpha'}^z
       \Big\rangle
        -
       \sum_{\alpha\alpha'}
       \Big\langle
            \Sigma_{\alpha}^z
            \Sigma_{\alpha'}^z
       \Big\rangle_0
  \bigg\}.
  \label{chi-perp-def}
\end{eqnarray}
The total Hamiltonian satisfies the SU(12) symmetry, so that
we can apply such a unitary transformation that make the spin
operators $S^z$ and $\Sigma_{\alpha}^z$ become to be diagonal.
This unitary transformation does not change the thermal average
of the spin operators, so that $\chi_{\perp}$ (\ref{chi-perp-def}) is,
\begin{eqnarray}
  \chi_{\perp} &=&
  \frac{g_{\rm{spin}}^2\mu_B^2}{T}
  \sum_{\lambda\lambda'}
  \frac{\sigma\sigma'}{4}
  \bigg\{
       \Big\langle
           X^{\lambda\lambda}
           \delta_{\lambda \lambda'}+
  \nonumber \\ &+&
           2\sum_{\alpha k}
           c_{\alpha k \lambda}^{\dag}
           c_{\alpha k \lambda}
           X^{\lambda'\lambda'}
  \nonumber \\ &+&
           \sum_{\alpha\alpha' k k'}
           c_{\alpha k \lambda}^{\dag}
           c_{\alpha k \lambda}
           c_{\alpha' k' \lambda'}^{\dag}
           c_{\alpha' k' \lambda'}
       \Big\rangle-
  \nonumber \\ &-&
       \sum_{\alpha \alpha' k k'}
       \Big\langle
           c_{\alpha k \lambda}^{\dag}
           c_{\alpha k \lambda}
           c_{\alpha' k' \lambda'}^{\dag}
           c_{\alpha' k' \lambda'}
       \Big\rangle_0
  \bigg\}.
  \label{chi-perp-sum}
\end{eqnarray}
We will estimate each term in the right hand side of eq.
(\ref{chi-perp-sum}) in turn. The first term gives,
\begin{subequations}
\begin{eqnarray}
&&  X_{dd} =
  \sum_{\lambda}
  \frac{\sigma^2}{4}
  \big\langle
      X^{\lambda\lambda}
  \big\rangle =
  \frac{1}{4},
  \label{X-d-d-derive}
\end{eqnarray}
where $\langle{X}^{\lambda\lambda}\rangle=\frac{1}{N}$ [$N=12$] does not
depend on the quantum number $\lambda$.

The second term in the right hand side of eq. (\ref{chi-perp-sum})
is,
\begin{eqnarray*}
  X_{d\alpha} &=&
  \sum_{k \lambda \lambda'}
  \frac{\sigma\sigma'}{2}
  \Big\langle
      c_{\alpha k \lambda}^{\dag}
      c_{\alpha k \lambda}
      X^{\lambda'\lambda'}
  \Big\rangle.
\end{eqnarray*}
The antiferromagnetic Kondo interaction makes the difference
between the two-particle states with parallel and antiparallel
states, therefore $X_{d\alpha}$ is not zero. In  subsection
\ref{subsec-proof-Puu-Pud} it is a proof that
\begin{eqnarray}
  P_{1}=
  \Big\langle
      c_{\alpha k \lambda}^{\dag}
      c_{\alpha k \lambda}
      X^{\lambda\lambda}
  \Big\rangle
  \label{P1-def}
\end{eqnarray}
does not depend on $\lambda$, whereas
\begin{eqnarray}
  P_{2}=
  \Big\langle
      c_{\alpha k \lambda}^{\dag}
      c_{\alpha k \lambda}
      X^{\lambda'\lambda'}
  \Big\rangle
  \label{P2-def}
\end{eqnarray}
does not depend on $\lambda$ and $\lambda'$ (just we should keep
$\lambda\ne\lambda'$). Using these equalities, we can write,
\begin{eqnarray}
  X_{d\alpha} &=&
  \frac{N}{2}
  \Big\{
      P_1-
      P_2
  \Big\}.
  \label{X-d-alpha-derive}
\end{eqnarray}

The third term in the right hand side of eq. (\ref{chi-perp-sum})
is,
\begin{eqnarray*}
  X_{\alpha\alpha'} &=&
  \sum_{k k' \lambda \lambda'}
  \frac{\sigma\sigma'}{4}
  \Big\langle
      c_{\alpha k \lambda}^{\dag}
      c_{\alpha k \lambda}
      c_{\alpha'k'\lambda'}^{\dag}
      c_{\alpha'k'\lambda'}
  \Big\rangle.
\end{eqnarray*}
$X_{\alpha\alpha'}$ can be estimated similarly to $X_{d\alpha}$.
The exchange interaction between the leads and the dot generates
an effective interaction between electrons in the leads. As a
result, the expectation value
$\langle{c}_{\alpha{k}\lambda}^{\dag}c_{\alpha{k}\lambda}%
c_{\alpha'k'\lambda'}^{\dag}c_{\alpha'k'\lambda'}\rangle$ depends
either $\lambda$ is equal to $\lambda'$ or not. Defining
$K_{1\alpha\alpha'}$ and $K_{2\alpha\alpha'}$,
\begin{eqnarray}
&&  K_{1\alpha\alpha'} =
  \sum_{kk'}
  \Big\langle
      c_{\alpha k \lambda}^{\dag}
      c_{\alpha k \lambda}
      c_{\alpha'k \lambda}^{\dag}
      c_{\alpha'k \lambda}
  \Big\rangle
  \nonumber \\
  && -
  \sum_{kk'}
  \Big\langle
      c_{\alpha k \lambda}^{\dag}
      c_{\alpha k \lambda}
      c_{\alpha'k \lambda}^{\dag}
      c_{\alpha'k \lambda}
  \Big\rangle_0,
  \label{K1-def}
\nonumber  \\
&&  K_{2\alpha\alpha'} =
  \sum_{kk'}
  \Big\langle
      c_{\alpha k \lambda}^{\dag}
      c_{\alpha k \lambda}
      c_{\alpha'k'\lambda'}^{\dag}
      c_{\alpha'k'\lambda'}
  \Big\rangle
  \nonumber \\
 && -
  \sum_{kk'}
  \Big\langle
      c_{\alpha k \lambda}^{\dag}
      c_{\alpha k \lambda}
      c_{\alpha'k'\lambda'}^{\dag}
      c_{\alpha'k'\lambda'}
  \Big\rangle_0,
  \ \ \
  \lambda\ne\lambda', \nonumber \\
&&  \label{K2-def}
\end{eqnarray}
($K_{1\alpha\alpha'}$ and $K_{2\alpha\alpha'}$ do not depend on
$\lambda$'s\footnote{Proof of the statement that
$K_{1\alpha\alpha'}$ and $K_{2\alpha\alpha'}$ do not depend on
$\lambda$'s is similar to the proof that $P_1$ and $P_2$ do not
depend on $\lambda$'s, subsection \ref{subsec-proof-Puu-Pud}.}), we
get
\begin{eqnarray}
  X_{\alpha\alpha'} &=&
  \frac{N}{4}
  \Big\{
      K_{1\alpha\alpha'}-
      K_{2\alpha\alpha'}
  \Big\}.
  \label{X-alpha-alpha-derive}
\end{eqnarray}
  \label{subeqs-X-derive}
\end{subequations}

With equations (\ref{X-d-d-derive}), (\ref{X-d-alpha-derive}) and
(\ref{X-alpha-alpha-derive}), the susceptibility $\chi_{\perp}$
takes the form
\begin{eqnarray}
  \chi_{\perp} &=&
  \frac{g_{\rm{spin}}^2\mu_B^2}{4T}
  \Big\{
      1+
      4N
      \big(
          P_1-P_2
      \big)+
  \nonumber \\ && +
      N
      \sum_{\alpha\alpha'}
      \big(
          K_{1\alpha\alpha'}-
          K_{2\alpha\alpha'}
      \big)
  \Big\}.
  \label{chi-perp-res}
\end{eqnarray}

Now  consider $\chi_{\parallel}$,
\begin{eqnarray}
  \chi_{\parallel} &=&
  \frac{g_{\rm{spin}}^2\mu_B^2}{T}
  \bigg\{
       \Big\langle
            S^xS^x+
            2\sum_{\alpha}
            S^x\Sigma_{\alpha}^x+
  \nonumber \\ && +
            \sum_{\alpha\alpha'}
            \Sigma_{\alpha}^x
            \Sigma_{\alpha'}^x
       \Big\rangle
       -
       \sum_{\alpha\alpha'}
       \Big\langle
            \Sigma_{\alpha}^x
            \Sigma_{\alpha'}^x
       \Big\rangle_0
  \bigg\}+
  \nonumber \\ &+&
  \frac{g_{\rm{spin}}g_{\rm{orb}}\mu_B^2}{T}
  \bigg\{
       \Big\langle
            S^xL^x+
            \sum_{\alpha}
            \Big(
            S^x\Lambda_{\alpha}^x+
            L^x\Sigma_{\alpha}^x
            \Big)+
  \nonumber \\ && +
            \sum_{\alpha\alpha'}
            \Sigma_{\alpha}^x
            \Lambda_{\alpha'}^x
       \Big\rangle -
       \sum_{\alpha\alpha'}
       \Big\langle
            \Sigma_{\alpha}^x
            \Lambda_{\alpha'}^x
       \Big\rangle_0
  \bigg\}+
  \nonumber \\ &+&
  \frac{g_{\rm{orb}}^2\mu_B^2}{T}
  \bigg\{
       \Big\langle
            L^xL^x +
            2\sum_{\alpha}
            L^x\Lambda_{\alpha}^x+
  \nonumber \\ && +
            \sum_{\alpha\alpha'}
            \Lambda_{\alpha}^x
            \Lambda_{\alpha'}^x
       \Big\rangle -
       \sum_{\alpha\alpha'}
       \Big\langle
            \Lambda_{\alpha}^x
            \Lambda_{\alpha'}^x
       \Big\rangle_0
  \bigg\}.
  \label{chi-par-def}
\end{eqnarray}
The right hand side of eq. (\ref{chi-par-def}) consists of three
blocks of terms consisting of the spin-spin, spin-orbital and
orbital-orbital correlation functions. The Kondo Hamiltonian
(\ref{HK-SU12-def}) does not contain the spin-orbital
interactions, so that the spin-orbital correlation functions are
zero. In order to derive the spin-spin part of $\chi_{\parallel}$,
we apply the unitary transformations to make the spins $S^x$ and
$\Sigma^x$ diagonal. It is easy to see that the spin part of
$\chi_{\parallel}$ gives eq. (\ref{chi-perp-res}). Consider now
the last block of terms coming from the orbital-orbital
correlations. Applying the unitary transformations to make the
orbital moments $L^x$ and $\Lambda_{\alpha}^x$ diagonal, we can
write the orbital moment contribution to $\chi_{\parallel}$ as,
\begin{eqnarray}
  \chi_{\parallel}^{\rm{orb}} &=&
  \frac{g_{\rm{orb}}^2\mu_B^2}{T}
  \sum_{mm'}
  m m'
  \bigg\{
       \Big\langle
            X^{mm}\delta_{mm'}\nonumber \\
  &&
            +
            2\sum_{\alpha k}
            c_{\alpha k \lambda}^{\dag}
            c_{\alpha k \lambda}
            X^{\lambda'\lambda'} \nonumber \\
            &&+
            \sum_{\alpha\alpha' k k'}
            c_{\alpha k \lambda}^{\dag}
            c_{\alpha k \lambda}
            c_{\alpha'k' \lambda'}^{\dag}
            c_{\alpha'k'\lambda'}
       \Big\rangle-
  \nonumber \\ &-&
       \sum_{\alpha\alpha' k k'}
       \Big\langle
            c_{\alpha k \lambda}^{\dag}
            c_{\alpha k \lambda}
            c_{\alpha'k' \lambda'}^{\dag}
            c_{\alpha'k'\lambda'}
       \Big\rangle_0
  \bigg\}.
  \nonumber \\
  \label{chi-par-orb}
\end{eqnarray}
The right hand side of eq. (\ref{chi-par-orb}) consists of the
terms coming from the dot-dot, dot-lead and lead-lead
correlations. We will consider all of them in turn.

The dot-dot correlation is,
\begin{subequations}
\begin{eqnarray}
  O_{dd} &=&
  \sum_{\lambda}
  m^2
  \big\langle
      X^{\lambda\lambda}
  \big\rangle =
  \frac{2}{3} =
  \frac{8}{3}~
  X_{dd}.
  \label{O-d-d-derive}
\end{eqnarray}

The dot-lead correlation is,
\begin{eqnarray*}
  O_{d\alpha} &=&
  2\sum_{k \lambda\lambda'}
  m m'
  \Big\langle
      c_{\alpha k \lambda}^{\dag}
      c_{\alpha k \lambda}
      X^{\lambda'\lambda'}
  \Big\rangle.
\end{eqnarray*}
Similarly to $X_{d\alpha}$, $O_{d\alpha}$ can be expressed in
terms of $P_1$ and $P_2$, eqs. (\ref{P1-def}) and (\ref{P2-def}),
as
\begin{eqnarray}
  O_{d\alpha} &=&
  16
  \big(
      P_1-
      P_2
  \big) =
  \frac{8}{3}~
  X_{d\alpha}.
  \label{O-d-alpha-derive}
\end{eqnarray}

Finally, the lead-lead correlation gives,
\begin{eqnarray*}
  O_{\alpha\alpha'} &=&
  \sum_{k k' \lambda\lambda'}
  m m'
  \Big\langle
      c_{\alpha k \lambda}^{\dag}
      c_{\alpha k \lambda}
      c_{\alpha'k'\lambda'}^{\dag}
      c_{\alpha'k'\lambda'}
  \Big\rangle.
\end{eqnarray*}
Similarly to $X_{\alpha\alpha'}$, $O_{\alpha\alpha'}$ can be
expressed in terms of $K_{1\alpha\alpha'}$ and
$K_{2\alpha\alpha'}$, eqs. (\ref{K1-def}) and (\ref{K2-def}),
as
\begin{eqnarray}
  O_{\alpha\alpha'} &=&
  8
  \big(
      K_{1\alpha\alpha'}-
      K_{2\alpha\alpha'}
  \big) =
  \frac{8}{3}~
  X_{\alpha\alpha'}.
  \label{O-alpha-alpha-derive}
\end{eqnarray}
  \label{subeqs-O-derive}
\end{subequations}

Combining equations (\ref{chi-par-orb}) and
(\ref{subeqs-O-derive}), we get $\chi_{\parallel}$ in the form,
\begin{eqnarray}
  \chi_{\parallel} &=&
  \frac{\mu_B^2}{T}~
  \bigg(
      \frac{g_{\rm{spin}}^2}{4}+
      \frac{2g_{\rm{spin}}^2}{3}
  \bigg)
  \times \nonumber \\ && \times
  \Big\{
      1+
      4N
      \big(
          P_1-P_2
      \big)+
  \nonumber \\ && +
      N
      \sum_{\alpha\alpha'}
      \big(
          K_{1\alpha\alpha'}-
          K_{2\alpha\alpha'}
      \big)
  \Big\}.
  \label{chi-par-res}
\end{eqnarray}
\textbf{Eqs. (\ref{chi-par-res}) and (\ref{chi-perp-res}) prove
eq. (\ref{ratio-suscept-proof}).}

\subsection{Proof of equations (\ref{P1-def}) and (\ref{P2-def})}
  \label{subsec-proof-Puu-Pud}

In order to prove eq. (\ref{P1-def}), we prove that two expected
values, $G_{\lambda\lambda}$ and $G_{\lambda'\lambda'}$, are equal
one to another,
\begin{eqnarray}
  &&
  G_{\lambda\lambda} =
  \Big\langle
      c_{\alpha k \lambda}^{\dag}
      c_{\alpha k \lambda}
      X^{\lambda\lambda}
  \Big\rangle,
  \nonumber \\ &&
  G_{\lambda'\lambda'} =
  \Big\langle
      c_{\alpha k \lambda'}^{\dag}
      c_{\alpha k \lambda'}
      X^{\lambda'\lambda'}
  \Big\rangle,
  \label{G-diag-def}
\end{eqnarray}
where $\lambda\ne\lambda'$. Let, for the brevity, enumerate the
quantum states of the quantum dot in such a way that $\lambda=1$
and $\lambda'=2$. The expected value $G_{\lambda'\lambda'}$
is invariant with respect to any unitary transformations,
$$
  c_{\alpha k \lambda'}
  \to
  \sum_{\lambda''}
  U_{\lambda'\lambda''}
  c_{\alpha k \lambda''},
  \ \ \
  c_{\alpha k \lambda'}^{\dag}
  \to
  \sum_{\lambda''}
  c_{\alpha k \lambda''}^{\dag}
  U_{\lambda''\lambda'},
$$
$$
  X^{\lambda'\lambda'}
  \to
  \sum_{\lambda''\lambda'''}
  U^{\lambda'\lambda''}
  X^{\lambda''\lambda'''}
  U^{\lambda'''\lambda'},
$$
where $U_{\lambda'\lambda''}$ is a unitary ${N}\times{N}$ matrix.
In particular, it is invariant with respect to the transformation given
by the matrix,
$$
  U =
  \left(
  \begin{array}{ccc}
  0 & 1 & 0
  \\
  1 & 0 & 0
  \\
  0 & 0 & \hat{I}_{10}
  \end{array}
  \right),
$$
where $\hat{I}_{10}$ is the $10\times10$ identity matrix. Applying this
transformation to the expected value $G_{\lambda'\lambda'}$,
we get the expected value $G_{\lambda\lambda}$, so that
$G_{\lambda\lambda}=G_{\lambda'\lambda'}$.

In order to prove eq. (\ref{P2-def}), we consider two expected values,
$G_{\lambda\lambda'}$ and $G_{\lambda\lambda''}$, are equal one
to another,
\begin{eqnarray}
  &&
  G_{\lambda\lambda'} =
  \Big\langle
      c_{\alpha k \lambda}^{\dag}
      c_{\alpha k \lambda}
      X^{\lambda'\lambda'}
  \Big\rangle, \nonumber \\
  &&
  G_{\lambda\lambda''} =
  \Big\langle
      c_{\alpha k \lambda}^{\dag}
      c_{\alpha k \lambda}
      X^{\lambda''\lambda''}
  \Big\rangle,
  \label{G-nondiag-def}
\end{eqnarray}
where $\lambda\ne\lambda'$, $\lambda\ne\lambda''$ and
$\lambda'\ne\lambda''$. Let, for the brevity, enumerate the
quantum states of the quantum dot in such a way that $\lambda=1$,
$\lambda'=2$ and $\lambda''=3$. The expected value
$G_{\lambda\lambda''}$ is invariant with respect to any unitary
transformations,
$$
  c_{\alpha k \lambda}
  \to
  \sum_{\lambda_1}
  U_{\lambda\lambda_1}
  c_{\alpha k \lambda_1},
  \ \ \
  c_{\alpha k \lambda}^{\dag}
  \to
  \sum_{\lambda_1}
  c_{\alpha k \lambda_1}^{\dag}
  U_{\lambda_1\lambda},$$
  $$
  X^{\lambda''\lambda''}
  \to
  \sum_{\lambda_1\lambda_2}
  U^{\lambda''\lambda_1}
  X^{\lambda_1\lambda_2}
  U^{\lambda_2\lambda''},
$$
where $U_{\lambda'\lambda''}$ is a unitary ${N}\times{N}$ matrix.
In particular, it is invariant with respect to the transformation given
by the matrix,
$$
  U =
  \left(
  \begin{array}{cccc}
  1 & 0 & 0 & 0
  \\
  0 & 0 & 1 & 0
  \\
  0 & 1 & 0 & 0
  \\
  0 & 0 & 0 & \hat{I}_{9}
  \end{array}
  \right),
$$
where $\hat{I}_{9}$ is the $9\times9$ identity matrix. Applying this
transformation to the expected value $G_{\lambda\lambda''}$,
we get the expected value $G_{\lambda\lambda'}$, so that
$G_{\lambda\lambda''}=G_{\lambda\lambda'}$.

\vspace{0.7in}
%%%%%%%%%%%%%%%%%%%%%%%%%

\end{document}